\documentclass[a4paper,11pt]{article}
\pdfoutput=1 

\usepackage{jheppub} 
                     
\usepackage{array,multirow}

\usepackage[T1]{fontenc} 
\usepackage{etoolbox}
\BeforeBeginEnvironment{tabular}{\begin{center}\scriptsize}
\AfterEndEnvironment{tabular}{\end{center}}
\usepackage{booktabs}
\usepackage{float}
\usepackage{scalerel}
\title{\boldmath Extraction of $|V_{cb}|$ from $B\to D^{(*)}\ell\nu_\ell$ and the Standard Model predictions of 
$R(D^{(*)})$}

\author{Sneha Jaiswal,Soumitra Nandi}
\author{and Sunando Kumar Patra}
 \affiliation{Indian Institute of Technology, North Guwahati, Guwahati 781039, Assam, India}

\emailAdd{sneha.jaiswal@iitg.ernet.in}
\emailAdd{soumitra.nandi@iitg.ernet.in}
\emailAdd{sunando.patra@gmail.com}

\usepackage{amsmath}
\usepackage{amssymb}
\usepackage{bm}
\usepackage{dcolumn,multirow,subfig}
\usepackage[normalem]{ulem}

\usepackage{array}

\usepackage{graphicx}
\usepackage{hyperref}

\newcommand{\ba}{\begin{array}}
\newcommand{\ea}{\end{array}}
\def\beq{\begin{equation}}
\def\eeq{\end{equation}}
\def\bea{\begin{eqnarray}}
\def\eea{\end{eqnarray}}
\def\nn{\nonumber}

\def\roughly#1{\mathrel{\raise.3ex\hbox
{$#1$\kern-.75em\lower1ex\hbox{$\sim$}}}}

\def\gsim{\roughly>}
\def\sla#1{\raise.15ex\hbox{$/$}\kern-.57em #1}

\def\order{\lower 1.8ex \hbox{\LARGE\~{}}}

\def\bdstlnu{$B \to D^* \ell \nu_{\ell}$ }
\def\bdlnu{$B \to D \ell \nu_{\ell}$ }
\def\errvec#1{$\left(\begin{smallmatrix} #1 \end{smallmatrix}\right)$}
\usepackage[normalem]{ulem}
\definecolor{darkgreen}{cmyk}{1,0,1,0.4}

\definecolor{pink}{cmyk}{0.4,1,0.3,0}

\def\com2#1{\textcolor{red}{\it{#1}}}

\abstract{We extract $|V_{cb}|$ from the available data in the decay $B \to D^{(*)}\ell\nu_{\ell}$.
Our analysis uses the $q^2(w)$ binned differential decay rates in different subsamples of $B\to D\ell\nu_\ell$ ($\ell = e, \mu$), while for the decay 
$B\to D^*\ell\nu_\ell$, the unfolded binned differential decay rates of four kinematic variables including the $q^2$ bins have been used.
In the CLN and BGL parameterizations of the form factors, the combined fit 
to all the available data along with their correlations yields $|V_{cb}| = (39.77 \pm 0.89)\times 10^{-3}$ and 
$(40.90 \pm 0.94)\times 10^{-3}$ respectively. In these fits, we have used the inputs from lattice and light cone sum rule (LCSR) along with the data. 
Using our fit results and the HQET relations (with the known corrections included) amongst the form factors, and parameterizing the unknown higher order corrections 
(in the ratios of HQET form factors) with a conservative estimate of the normalizing parameters, we obtain $R(D^{*}) = 0.259 \pm 0.006$ (CLN) and 
$R(D^*) = 0.257 \pm 0.005$ (BGL). }

\begin{document} 
\maketitle
\flushbottom

\section{Introduction}

One of the primary goals of the study of $B$ meson decays and mixing is to construct the unitarity triangle (UT). 
In this regard, the CKM elements $V_{ub}$ and $V_{cb}$ play an important role. 
Hence, precise measurements of these elements are of utmost importance.   

The tree level semileptonic decays $b\to c\ell\nu_{\ell}$ ($\ell = e, \mu$) are crucial for the determination of $|V_{cb}|$. 
It can be extracted from both exclusive decays, like $B\to D^{(\ast)}\ell\nu$, and inclusive decays, like 
$B \to X_c \ell\nu_{\ell}$. The inclusive channels are relatively clean, and the 
decay rates have a solid description via operator product expansion (OPE) or heavy quark expansion (HQE) \cite{OPE}. 
The exclusive semileptonic decays have similar solid descriptions in terms of heavy quark effective theory 
(HQET) \cite{exclusive}.  Contrary to the inclusive decays, the non-perturbative unknowns in the exclusive decays 
can not be extracted experimentally. One needs to calculate them and that is where the major challenges lie. 
At the moment, the most precise determinations of $|V_{cb}|$ from inclusive \cite{inclusive} and exclusive decays 
\cite{hfag} differ from each other at $\approx 3\sigma$ confidence level (CL). Recently it has been shown that 
the Caprini-Lellouch-Neubert (CLN) \cite{CLN} and Boyd-Grinstein-Lebed (BGL) \cite{BGL} parameterizations lead to
different results for the exclusive determinations of $|V_{cb}|$ \cite{Bigi:2017}. In their analysis they have used 
up-to-date lattice calculations of the form factors along with the available experimental results from Belle 
\cite{Abdesselam:2017kjf}. 

Form factors, fitted from the decays $B\to D^{(\ast)}\ell\nu$, play a crucial role in the Standard Model (SM)
predictions of $R(D^{(*)}) = Br(B\to D^{(*)}\tau\nu_{\tau})/Br(B\to D^{(\ast)}\ell\nu)$. In the decays
$B\to D^{(*)}\tau\nu_{\tau}$, there are additional form factors that can not be extracted directly from the available 
data on $B\to D^{(\ast)}\ell\nu$. Therefore, one needs to rely on various theory inputs, the HQET relations between 
the form factors in particular. The SM predictions of $R(D^{(*)})$ using the CLN parametrization of the form factors and the inputs from lattice and 
HQET are given by \cite{Aoki:2016,Fajfer:2012vx}
\begin{equation}
 R(D) = 0.300 \pm 0.008, \ \ \ \ \ R(D^{*}) = 0.252 \pm 0.003.
\end{equation}
Recently, the SM prediction of $R(D)$ has been updated \cite{Bigi:2016mdz} using the lattice input on the form factors in 
$B\to D\ell\nu_{\ell}$ beyond the zero recoil \cite{Lattice:2015rga,Na:2015kha}. The updated value is 
$R(D) = 0.299 \pm 0.003$ \cite{Bigi:2016mdz}, which is the most precise estimate so far. Also, $R(D^{*})$ and $|V_{cb}|$
have been updated from a combined fit to the $B\to D^{(\ast)}\ell\nu$ differential rates and angular distributions, 
including  ${\cal O}(\Lambda_{QCD}/m_{b,c},\alpha_s)$ terms in HQET form factors \cite{Bernlochner:2017}. They have obtained 
\begin{equation}
 |V_{cb}| =(39.3 \pm 1.0) \times 10^{-3}, \ \ \ \ R(D^{*}) = 0.257 \pm 0.003
\end{equation}
using the lattice results on the form factors and the QCD sum rule (QCDSR) predictions \cite{Neubert:1992f,Neubert:1992s} for the 
HQET parameter as inputs in their analysis. The ratio of the form factors $R_0(w)$ \cite{CLN,Fajfer:2012vx}, where $w$ 
is the recoil angle between $B$ and $D^*$, and its value at zero recoil, $R_0(1)$, play a crucial role in the determination 
of $R(D^{*})$. The estimate of $R_0(1)$ depends on the HQET parameter $\eta(1)$, for which the QCDSR prediction is available \cite{Bernlochner:2017}. 
However, it is also possible to fit it directly from the available experimental data and the lattice inputs.  
As an example, we can see the Table-II of ref. \cite{Bernlochner:2017}, and note that the fit results for $\eta(1)$ (with lattice as input) deviates 
from that predicted in QCDSR by more than 1$\sigma$.  Also, the ratio of the form factors in $B\to D^{(*)}\ell\nu_{\ell} $ differ from that predicted by 
lattice \cite{Bigi:2016mdz,talk}. Therefore, the legitimate query is whether this is due to the missing pieces in the HQET relations between the 
form factors (i.e. corrections at order $\alpha_s^2$ and $\Lambda_{QCD}^2/m_{b,c}^2$).

In this article, we have extracted $|V_{cb}|$ independently from the fits to the available data on the differential  
rates and angular distributions in $B\to D\ell \nu_{\ell}$ and $B\to D^{*}\ell \nu_{\ell}$, using the CLN and BGL parameterization of the form factors.
We have then performed a combined analysis of the complete data set in both the parameterizations of the form factors and extracted $|V_{cb}|$, $R(D)$ and $R(D^{*})$.
In this analysis, along with the experimental data, we use the lattice predictions for the form factors as inputs \cite{Lattice:2015rga,Na:2015kha}. 

As mentioned earlier, we have an additional form factor in $R(D^{*})$, that cannot be constrained from experimental data alone and we need additional theory 
inputs. In order to predict $R(D^{*})$, we define the HQET relations between the form factors with the known corrections \cite{CLN,Bernlochner:2017}, which 
are represented in terms of the sub-leading Isgur-Wise functions. We constrain those functions (HQET parameters) from a 
fit to the ratios of the form factors used in our analysis, and the synthetic data for these ratios are obtained using directly our fit results or from lattice and 
light cone sum rule (LCSR). We repeat the analysis by considering additional parameters ($\Delta$) parameterizing the missing higher order corrections in the ratios 
of the HQET form factors, and have made a rough estimate of the probable size of these $\Delta$s in different ways with the available resources. We have 
considered the additional errors conservatively while predicting the SM value of $ R(D^{*})$.

\section{Inputs}\label{sec:results}


For \bdlnu data, we depend on the latest fully reconstructed measurement from Belle \cite{Glattauer:2015teq}, 
but instead of the combined result of 10 $w$ bins (in table II of that paper), we use the full dataset including 
all the four subsamples $B^+ \to \bar{D^0} e^+ \nu_e$, $B^+ \to \bar{D^0} \mu^+ \nu_{\mu}$, $B^0 \to D^- e^+ \nu_e$, 
and $B^0 \to D^- \mu^+ \nu_{\mu}$, with 40 data-points, along with their statistical and systematic uncertainties and the full 
systematic correlation matrix. These are available in \cite{belle:dlnusupp}. 
We also use the values of the form factors 
$f_+$ and $f_0$ at $w$ values $1$, $1.08$, and $1.16$ with the full covariance matrix supplied by MILC \cite{Lattice:2015rga}. 
On the other hand, the HPQCD collaboration uses BCL parametrization to present their results. While using the HPQCD 
results \cite{Na:2015kha}, we recognize (following the observation made by ref. \cite{Bigi:2016mdz}) that their simulations extend to 
a maximal value of $z=0.013$ ($w \approx 1.11$), and thus use synthetic data for $f_{+,0}$ at $w=1.00,~1.06,$ and $1.12$. These  are 
listed in table \ref{tab:fvalsHP}, with their uncertainties and correlation matrix. Belle, however, has used the same $w$ points 
as MILC to calculate HPQCD synthetic data in their analysis. 
We will explicitly mention our inputs whenever we are using them.

\begin{table}[t] 
 \begin{center}
  \def\arraystretch{1.2}
  \begin{tabular}{|cccccccc|}
\hline
$f_+(w)$ & Value from & \multicolumn{6}{c|}{Correlation} \\
    \& $f_0(w)$ & HPQCD & & & & & & \\
    \hline
     $f_+(1)$ & $1.178 (46)$ & 1. & 0.994 & 0.975 & 0.507 & 0.515 & 0.522 \\
     $f_+(1.06)$ & $1.105 (42)$ &  & 1. & 0.993 & 0.563 & 0.576 & 0.587 \\
     $f_+(1.12)$ & $1.037 (39)$ &  &  & 1. & 0.617 & 0.634 & 0.649 \\
     $f_0(1)$ & $0.902 (41)$ &  &  &  & 1. & 0.997 & 0.988 \\
     $f_0(1.06)$ & $0.870 (39)$ &  &  &  &  & 1. & 0.997 \\
     $f_0(1.12)$ & $0.840 (37)$ &  &  &  &  &  & 1. \\
    \hline
     & Value from & \multicolumn{6}{c|}{} \\
     & MILC & & & & & & \\
     \hline
     $f_+(1)$ & $1.1994 (95)$ & 1. & 0.967 & 0.881 & 0.829 & 0.853 & 0.803 \\
     $f_+(1.08)$ & $1.0941 (104)$ &  & 1. & 0.952 & 0.824 & 0.899 & 0.886 \\
     $f_+(1.16)$ & $1.0047 (123)$ &  &  & 1. & 0.789 & 0.890 & 0.953 \\
     $f_0(1)$ & $0.9026 (72)$ &  &  &  & 1. & 0.965 & 0.868 \\
     $f_0(1.08)$ & $0.8609 (77)$ &  &  &  &  & 1. & 0.952 \\
     $f_0(1.16)$ & $0.8254 (94)$ &  &  &  &  &  & 1. \\     
 \hline    
\end{tabular}
\end{center}
\caption{Lattice QCD results of $f_+$ and $f_0$ for different values of $w$. The upper half of the table have been obtained using the fit results from the 
HPQCD collaboration \cite{Na:2015kha}, and the lower half are the results obtained by the Fermilab MILC collaboration \cite{Lattice:2015rga}.}
\label{tab:fvalsHP}
\end{table}

\begin{table}[!hbt] 
 \begin{center}
  \begin{tabular}{|c c|}
    \hline
    Source & $\mathcal{G}(1)$ \\
    \hline
    Fermilab/MILC \cite{Lattice:2015rga} & $1.0541(83)$ \\
    HPQCD \cite{Na:2015kha} & $1.035(40)$ \\
    HQE(BPS Expansion) \cite{Uraltsev:2003ye} & 1.04(2) \\
    \hline
  \end{tabular}
 \end{center}
 \caption{Different values of $\mathcal{G}(1)$ used in \bdlnu fits.}
 \label{tab:g1vals}
\end{table}

In addition to using the dispersions relations, CLN parametrization \cite{CLN} uses Heavy Quark Effective Theory 
(HQET) to strengthen the unitarity bounds and as a consequence this establishes approximate relations between the slope and 
the higher power coefficients of the form factors (valid within $\approx 2\%$). Other than $|V_{cb}|$, only two parameters parametrize 
the form factors under this scenario: $\rho_D^2$ and $\mathcal{G}(1)$. The form factor normalization $\mathcal{G}(1)$ is predicted by 
both HPQCD and Fermilab/MILC. There is one HQE result based on Bogomol'nyi-Prasad-Sommerfield (BPS) symmetry (partially)\cite{Uraltsev:2003ye} as well. 
These are listed in table \ref{tab:g1vals} and $\mathcal{G}(1)$ is used as a nuisance parameter in some of our fits.

In our analysis, for the \bdstlnu data, we mainly depend on the unfolded binned differential decay rates by Belle. For four 
kinematic variables $w,~\cos{\theta_v},~\cos{\theta_l}$ and $\chi$, with 10 bins each, this amounts to a total of 40 data points, 
their uncertainties and the full correlation matrix \cite{Abdesselam:2017kjf}. Other than these, we make use of the zero-recoil 
value of the form factor $h_{A_1}(w)$ from unquenched Fermilab/MILC lattice data \cite{Bailey:2014tva}:
\beq\label{ha11latval}
h_{A_1}(1) = 0.906 \pm 0.013\,.
\eeq
In addition to these, we have used, in few cases, the inputs from light cone sum rule (LCSR) \cite{Faller:2008}: 
\begin{align}\label{lcsr}
 h_{A_1}(w_{max} ) &= 0.65 (18) , \ \ \ R_1(w_{max}) = 1.32 (4),\ \ \  R_2(w_{max}) = 0.91 (17), 
\end{align}

and the following inputs throughout our analysis:
\begin{align}\label{thinputs}
\bar{m}_b(\bar{m}_b)&=4.163\pm 0.016\ {\it GeV}, \ m_c(3 GeV)=0.986\pm 0.013\ {\it GeV}, \nn \\ 
&{} \alpha_S(\bar{m}_b(\bar{m}_b))=0.2268\pm 0.0023. 
\end{align}

\section{CLN parameterization : Fit results}
\begin{table}[!hbt] 
 \begin{center}
  \def\arraystretch{1.2}
  \begin{tabular}{|c|c|c|c|c|}
     \hline
     Constraints & $|V_{cb}|$  & $\chi^2_{min} / d.o.f$ & $p$-value & $R(D)$\\
   & ($\times 10^3$) & &  ($\%$) & \\
     \hline
    Using only $\mathcal{G}(1)$ & & &  & \\
      \textbf{HPQCD+MILC} & $\bm{39.97(1.34)}$ & {\bf 23.04/39} & {\bf 98.02} & {\bf 0.299(6)} \\
      \textbf{HPQCD+MILC+BPS} & $\bm{40.04(1.33)}$ & \textbf{23.42/40} & \textbf{98.30} & \textbf{0.299(6)} \\
    \hline
      Belle \cite{Glattauer:2015teq} & $39.86(1.33)$ & 4.57/8 & 80 & 0.298(6) \\
    \hline
 Using only $f_+(w)$ &   &    &   &   \\
 {\bf HPQCD + MILC} &  {\bf 40.84(1.15)} & {\bf 31.22/43} & {\bf 90.91} &  ${\bf 0.305(3)}$ \\  
\hline 
 \end{tabular}
 \end{center}
 \caption{Result of the fit to the experimental data in \bdlnu using only $\mathcal{G}(1)$ (first two rows), and using $f_+(w)$ ($w=1$ and $w\ne 1$)
 from lattice (MILC and HPQCD listed in table  \ref{tab:fvalsHP}) with the CLN parametrization of the form factors.}
 \label{tab:clndlnumulti}
\end{table}

\subsection{Fit from \bdlnu data}

As shown in Table \ref{tab:clndlnumulti}, when we fit the available data using CLN parameterization for the form factors, we use different combinations of the predicted 
values of $\mathcal{G}(1)$. The best results are obtained when all the inputs are combined together, and the corresponding extracted values of $|V_{cb}|$ and 
$R(D)$ are shown in table \ref{tab:clndlnumulti}.
To do a preliminary cross-check of the validity of the fits, we have completely reproduced the table (V) of ref. \cite{Glattauer:2015teq}, except the last column, 
where the authors quote the fit results after averaging the separate samples. We have instead used the whole 
40-data-point-long sample with the full correlation matrix and have considered values of $m_e$ and $m_{\mu}$ to incorporate the correct values of 
$w_{max}$ consistent with experimental results. The reason for doing this is two-fold: (a) The increased number of degrees of freedom improves the quality of 
fit considerably. Even with $d. o. f$ increased by a factor of 5, $\chi^2_{min}/d. o .f$ hardly increases. However, as can be 
seen from the $p$-values for our results in table \ref{tab:clndlnumulti}, there is a considerable improvement in the goodness-of-fit.
(b) We wanted to use the full correlation in the data. The fact that our results match with Belle for all sub-samples separately
up to the second decimal place, while the full fit very slightly differs from the averaged result, makes the importance of considering the correlations even
more pertinent. Using $\mathcal{G}(1)$ from HPQCD and MILC with or without the constraint from BPS 
gives us our obtained result, given in bold-faced font.  For a comparison, in the third row of table \ref{tab:clndlnumulti}, we 
quote the experimental results too. The experimental analysis fits the quantity $\eta_{EW} \mathcal{G}(1) |V_{cb}|$ and then uses the MILC value of 
$\mathcal{G}(1)$ and the electroweak correction factor $\eta_{EW} = 1.0066$ to calculate $|V_{cb}|$. We note a little increase in the central values of our estimates 
of $|V_{cb}|$ with respect to that of Belle, however, the percentage error in the estimate does not change.  Also, as we have 
fitted $\mathcal{G}(1)$ separately under the above-mentioned constraints, it has a non-zero correlation with $|V_{cb}|$. With the increased 
number of data points, we obtain a better fit than \cite{Glattauer:2015teq}, as can be seen from the $p$-values. The extracted values of $R(D)$ are also consistent 
with that extracted in \cite{Glattauer:2015teq}. 

The last row of the same table represents the results obtained from a fit to the available 
experimental data along with the lattice inputs on $f_+(w)$ (table \ref{tab:fvalsHP}). We note that the central value of the fitted $V_{cb}$ is increased by 
$\approx$ 2\% while the percentage error has reduced from 3.3\% to 2.8\%. Also, now we can compare the predicted values of $R(D)$, which are obtained from the fit 
with and without the lattice inputs on $f_+(w)$. The central value of the predicted $R(D)$ has increased due to the use of $f_+(w)$, and there is a 
considerable reduction in the percentage error of the estimate. Our result is in agreement with the prediction of the earlier analysis \cite{Bigi:2016mdz}. In our fit,
we do not include the inputs on $f_0(w)$ from lattice, inclusion of which makes the fit worse (with a p-value $<$ 1\%). However, the fit is not that bad 
(p-value $\approx$ 55\%) if we drop all the available inputs from MILC and just use the inputs from HPQCD along with the experimantal data.

\subsection{Fit from \bdstlnu~ data}

\begin{table}[!t] 
 \begin{center}
 \def\arraystretch{1.3}
  \begin{tabular}{|c|c|c|}
     \hline
     &{Data+Lattice} & {Data+Lattice+LCSR} \\
     \hline
       Parameters/ & Best Fit $\pm$ Err. & Best Fit $\pm$ Err. \\
       Observables& Values  & Values   \\
      \hline
      $\bm{|V_{cb}| \times 10^{3}}$ & $\bm{38.23 \pm 1.46}$ &  $\bm{38.15 \pm 1.43}$   \\
      \hline
      $\rho_{D^*}^2$ & $1.17 \pm 0.15$ &  $1.16 \pm 0.14$ \\
      $R_1(1)$ & $1.39 \pm 0.09$ &  $1.37 \pm 0.04$ \\
      $R_2(1)$ & $0.91 \pm 0.08$ &  $0.91 \pm 0.07$ \\
      $h_{A_1}(1)$ & $0.91 \pm 0.01$ & $0.91 \pm 0.01$  \\
    \hline
  $\chi^2_{min} $ & 34.14 & 34.62 \\
 dof & 36 & 39  \\
 $p$-value & 55.73\% & 69.10\% \\
 \hline 
      $R_0(1)$ & $1.191 \pm 0.017 $ & $1.195 \pm 0.017 $ \\ 
     \hline
     $\bm{R(D^{*})}$ & $\bm{0.255 \pm 0.004}$ & $\bm{0.255 \pm 0.004}$ \\
 \hline
 \end{tabular}
 \end{center}
 \caption{Fit results with CLN for \bdstlnu, combined with constraint from eq. (\ref{ha11latval})}
 \label{tab:resFdstlnu}
\end{table}

\begin{table}[!t] 
 \begin{center}
 \def\arraystretch{1.3}
  \begin{tabular}{|c|c|c|}
     \hline
     &{Data+Lattice} & {Data+Lattice+LCSR} \\
     \hline
       Parameters & Best Fit $\pm$ Err. & Best Fit $\pm$ Err. \\
       & Values  & Values   \\
      \hline
      $\bm{|V_{cb}| \times 10^{3}}$ & $\bm{39.82 \pm 0.90}$ &  $\bm{39.77 \pm 0.89}$    \\
      \hline
      $\rho_{D}^2$ & $1.138 \pm 0.023$ &  $1.138 \pm 0.023$ \\
      $\mathcal{G}(1)$ & $1.058 \pm 0.007$ & $1.058 \pm 0.007$ \\
      $\rho_{D^*}^2$ & $1.269 \pm 0.123$ & $1.251 \pm 0.113$ \\
      $R_1(1)$ & $1.386 \pm 0.087$ & $1.371 \pm 0.036$ \\
      $R_2(1)$ & $0.880 \pm 0.073$ &  $0.888 \pm 0.065$\\
      $h_{A1}(1)$ & $0.900 \pm 0.012$ & $0.900 \pm 0.012$ \\
           \hline
  $\chi^2_{min} $ & 67.34 & 67.99 \\
 dof & 79 & 82  \\
 $p$-value & 82.21\% & { 86.66\%} \\
 \hline
 \end{tabular}
 \end{center}
 \caption{Results of the combined fit to the data in $B\to D^{(*)}\ell\nu_{\ell}$ with CLN.}
 \label{tab:resFcomb}
\end{table}

For the decay \bdstlnu, details of the parametrization of the form factors can be seen in refs. \cite{CLN,Fajfer:2012vx}.  
In addition to $|V_{cb}|$, there are 4 other parameters to fit in this case, of which $h_{A_1}(1)$ is put into the fit as a nuisance
parameter with input from eq. (\ref{ha11latval}). Fit results are listed in table \ref{tab:resFdstlnu}. $|V_{cb}|$ obtained from this 
fit has slightly larger uncertainty than that has been obtained from the \bdlnu fit, although there is a small decrease in the central value. The overall 
multiplicative parameter here is $h_{A_1}(1) |V_{cb}|$, so $h_{A_1}(1)$ has a correlation with $|V_{cb}|$ in our fit. Our fit 
values agree with those obtained in earlier analyses \cite{Abdesselam:2017kjf,Bigi:2017}, and in \cite{Bigi:2017jbd} \footnote{The preprint of a parallel work 
shares a similar publication timeline with our work.}. Results obtained from a similar kind of fit, where, in addition to lattice, inputs from LCSR (eq. \ref{lcsr})
have been incorporated, are shown in the right panel of table \ref{tab:resFdstlnu}. We note that although there are no considerable changes in the 
fitted values of $|V_{cb}|$, the error in the extracted value of $R_1(1)$ has reduced from 6.5\% to 3\%. The uncertainties in all the other fit parameters have 
reduced (though not considerably).

The calculation of $R(D)$ in CLN parametrization is straightforward. However, as mentioned earlier, calculation of $R(D^*)$ depends on an additional 
form factor ratio $R_0(w)$ and its value calculated at zero-recoil, which can not be fitted from $B\to D^*\ell\nu_{\ell}$.
The form factor ratios $R_i(w)$s are expressed as the ratios of the HQET form factors \cite{CLN,Bernlochner:2017} $h_i$s, like 
\begin{align}\label{Riws}
R_1(w) = \frac{h_{v}}{h_{A_1}} , \ \ \ \ \ \ \ R_2(w) = \frac{h_{A_3}}{h_{A_1}} + r_{D^*} \frac{h_{A_2}}{h_{A_1}}, \nn \\
R_0(w) = \frac{(w+1)}{(1 + r_{D^*})} - \frac{(w - r_{D^*})}{(1 + r_{D^*})} \frac{h_{A_3}}{h_{A_1}} - \frac{(1 - w r_{D^*})}{(1 + r_{D^*})}\frac{h_{A_2}}{h_{A_1}},
\end{align}
where $r_{D^*} = m_{D^*}/m_B$. The $h_i$s, and hence the form factor ratios include the corrections at order $\alpha_s$ and $\Lambda_{QCD}/m_{b,c}$. They are expressed   
in terms of a few sub-leading Isgur-Wise functions (HQET parameters), like $\eta(1),\eta^{\prime}(1),\chi_2(1),\chi^{\prime}_2(1)$, and 
$\chi^{\prime}_3(1)$. We note that both $R_2(w)$ and $R_0(w)$ are sensitive to the ratios $h_{A_3}/h_{A_1}$ and $h_{A_2}/h_{A_1}$. In the HQET, the $R_1(1)$, $R_2(1)$ 
and $R_0(1)$ are obtained from eq. \ref{Riws} by taking the limit $w\to 1$, and all of them are functions of the above mentioned HQET parameters. 
Hence, the $R_0(1)$ can be estimated only after the extractions of these HQET parameters.

Also, the form factor ratios $f_+(w)/f_0(w)$ can be expressed in terms of the ratios of the HQET form factors, like 
\begin{equation}
 \frac{f_+(w)}{f_0(w)} = \frac{(1 + r_D)^2}{2 r_D (w + 1)}\left(\frac{ \frac{h_-}{h_+}\frac{1 - r_D}{1 + r_D} - 1}
 {\frac{h_-}{h_+}\frac{1 + r_D}{1 - r_D}\frac{w - 1}{w + 1} - 1 }\right),
 \label{fpbyf0}
\end{equation}
with $r_D = m_D/m_{B}$. The HQET form factors $h_+$ and $h_-$ are also known at order $\alpha_s$ and $\Lambda_{QCD}/m_{b,c}$, and can be 
expressed in terms of the above mentioned five HQET parameters.
  
In the CLN parameterization of the form factors, we have expressed $R_i(w)$ as given in eq. (B7) of ref. \cite{Fajfer:2012vx} and fit 
$R_2(1)$ and $R_1(1)$ from the available data. Using these fit results and the inputs from lattice, we then estimate $R_0(1)$ after extracting the HQET parameters for 
the cases mentioned in table \ref{casescln}. The predictions of $R(D^{*})$ in both the cases are shown in table \ref{tab:resFdstlnu}.

\begin{table}[t]
 \begin{center}
  \begin{tabular}{|c|c|}
  \hline
Cases~~~  & Inputs for the fits \\
\hline
case-1~~~ & $R_1(1)$, $R_2(1)$, $f_+(w)/f_0(w)$     \\
 & for $w$=1, 1.08, 1.16 (MILC) \\
                                &   and $w$= 1.03, 1.06, 1.09, 1.12 (HPQCD) \\
\hline 
case-2~~~ & case-1 with $R_1(w_{max})$, $R_2(w_{max})$ from LCSR. \\
\hline        
\end{tabular}
\end{center}
\caption{Different cases for the fit of sub-leading Isgur-Wise functions. }
\label{casescln}
\end{table}

\subsection{Combined fit from \bdlnu and \bdstlnu data:}

\begin{figure*}[htbp]
\centering
\includegraphics[scale=0.4]{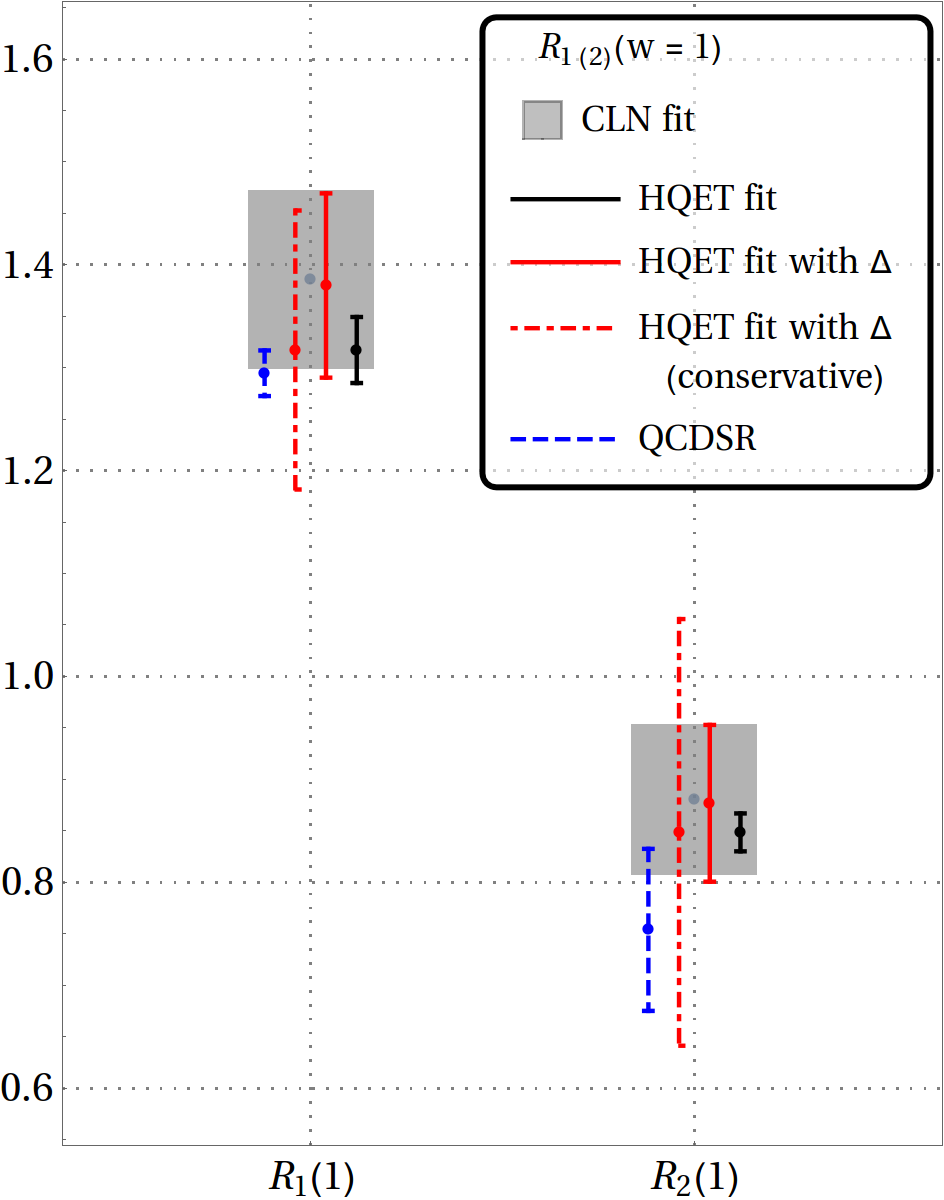}~~~~~~~~~~~~
\includegraphics[scale=0.4]{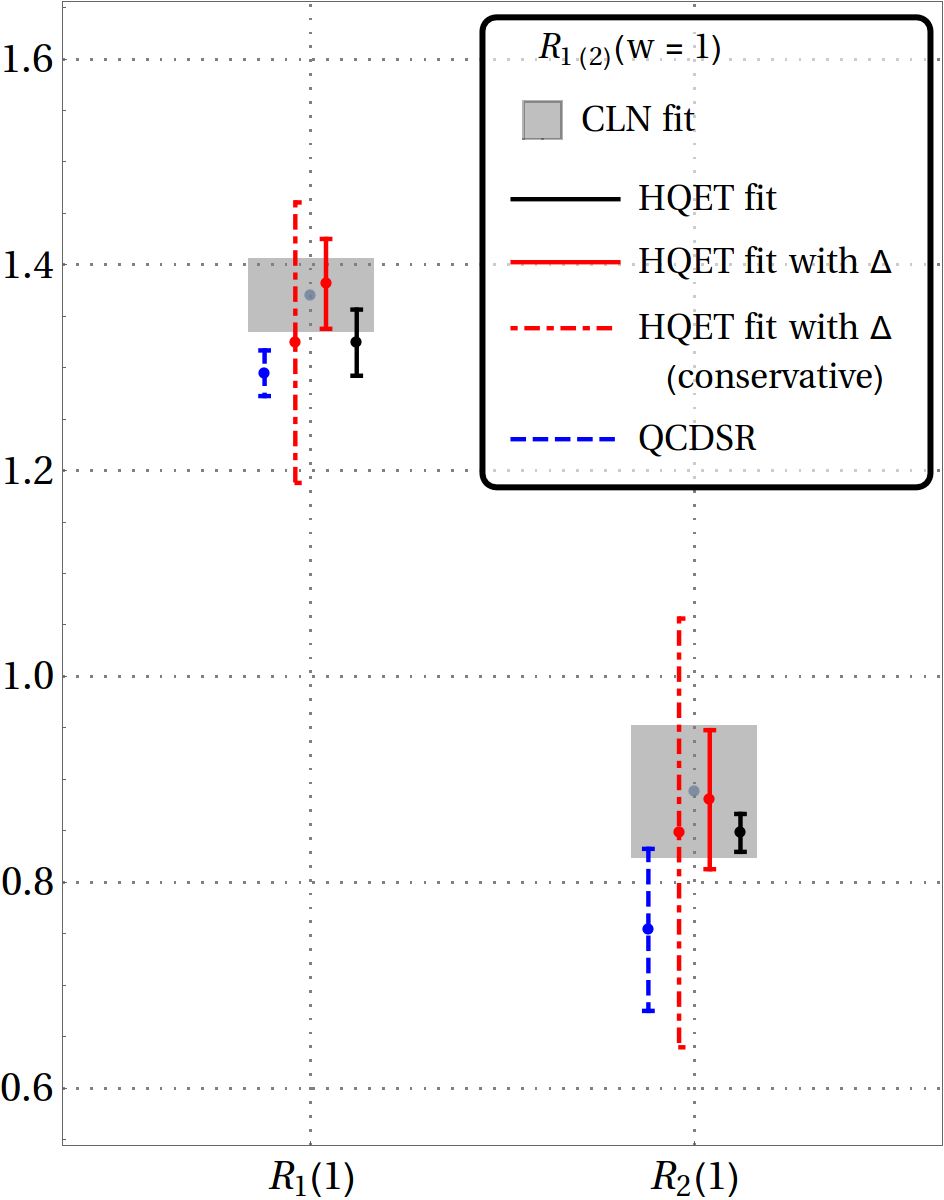}
\caption{Comparisions between different $R_1(1)$ and $R_2(1)$ which are obtained from different fits, with (right) and without (left) the inputs from 
LCSR, and QCDSR. For the conservative estimates of $R_{1(2)}(w=1)$, $\Delta_{v}$ is taken as $1 \pm 0.1$ and each of $\Delta_{21}$ and $\Delta_{31}$ 
is considered as $1 \pm 0.2$. }
\label{fig:R1R2}
\end{figure*}

\begin{table}[htbp]
\begin{center}
  \def\arraystretch{1.5}
  \begin{tabular}{|c | c | c |c |c| }
     \hline
     Para- & case-1    & case-1 & case-2 & case-2     \\
     meters &    & with $\Delta$s&     & with $\Delta$s  \\
     \hline
     $\eta(1)$         & 0.39(3) & 0.40(5)  & 0.39(3)  & 0.40(5)      \\
$\eta^{\prime}(1)$     & -0.002(100) & 0.004(101) & -0.03(9)  & 0.001(101)    \\
$\chi_2(1)$            & -0.08(1) & -0.06(1)& -0.08(1) & -0.06(1)     \\
${\chi_2}^{\prime}(1)$ & -0.003(2)& -0.003(2) &-0.001(2) & -0.002(2)  \\
${\chi_3}^{\prime}(1)$ & 0.04(2) & 0.04(2) & 0.05(2)  & 0.04(2)  \\
\hline
$\Delta_v$ & -  & 1.05(6)  & -  & 1.05(2)  \\  
$\Delta_{21}$ & - & 1.00(20)& -   & 1.00(20)    \\  
$\Delta_{31}$ & -  & 1.04(8) & -   & 1.04(7)        \\
$\Delta_{\mp}$ &    & 1.00(20)& -     & 1.00(20)   \\
      \hline     
$\chi^2_{min} $ & 4.34 & 3.36 & 9.20 & 3.62   \\
 dof           & 7      & 7    & 9    & 9          \\
 $p$-value & 73.95\% & 85.00\%  & 41.90\%  & 93.46\%  \\
 \hline
 \end{tabular}
 \end{center}
 \caption{The fit results for the subleading Isgur-Wise functions and the $\Delta$s (see text).}
 \label{tab:hqetcln}
\end{table}

Combining the full set of $w$-binned data from all subsamples for \bdlnu and the full data-set with all four variables 
for \bdstlnu~gives us the unique opportunity to not only simultaneously fit all the form factor parameters along 
with $|V_{cb}|$, but also predict the values and correlated uncertainties of $R(D)$ and $R(D^*)$ together.
The fit results are listed in table \ref{tab:resFcomb}.
The combined fit of the data in \bdlnu and \bdstlnu~shows considerable improvement over that obtained from the analysis of only the decay \bdstlnu.  Though changes in the fitted values of $R_1(1)$ and $R_2(1)$ are small, 
extracted uncertainties of $|V_{cb}|$ reduces to $\approx$ 2\% and the central value of $\rho^2_{D^*}$ increases by approximately 8\% in the combined analysis. 
Using the fit results given in table \ref{tab:resFcomb} and the inputs from lattice and LCSR, we obtain $R_0(1)$ and hence $R(D^*)$ for the cases given in 
table \ref{casescln}. For details, see 2nd (case-1) and 4th (case-2) columns of table \ref{tab:clnrdst}. Also in the combined analysis of all datasets, 
the central values of the predicted $R(D^*)$ increase by approximately 2\%, and the uncertainties are about 1\%.

\begin{table}[htbp]
\begin{center}
  \def\arraystretch{1.5}
  \begin{tabular}{|c | c | c |c |c| }
     \hline
     Param- & case-1    & case-1 & case-2 & case-2     \\
      eters/&  & with    &    &  with \\  
      Obser-&    & $\Delta_{31}$ \& $\Delta_{21}$ &     & $\Delta_{31}$ \& $\Delta_{21}$  \\
      vables&    &  = 1.00(20) &     &   = 1.00(20)  \\
\hline         
$R_0(1)$ & 1.192(17) & 1.192(101)  & 1.196(17) & 1.196(102) \\ 
 \hline
$\bm{R(D)}$ & \bf{0.304(3)}  &\bf{0.304(3)}  & \bf{0.304(3)} &  \bf{0.304(3)}                  \\
$\bm{R(D^{*})}$ & \bf{0.259(3)} & \bf{0.259(6)} & \bf{0.259(3)} &  \bf{0.259(6)}  \\
\hline
 Corr($R(D)$  & 0.21  &  0.12  &  0.20 &   0.11 \\
 -$R(D^*)$) &  &   & &   \\
 \hline
 \end{tabular}
 \end{center}
 \caption{The SM predictions for $R(D)$ and $R(D^*)$.}
 \label{tab:clnrdst}
\end{table}

\subsubsection{$R(D^*)$ with the additional error $\Delta$}

For completeness, as was mentioned in the introduction, we have also introduced additional parameters ($\Delta$), which parametrize the 
unknown higher order corrections in the ratios of the HQET form factors, such as $h_{v}/h_{A_1}$, $h_{A_3}/h_{A_1}$, $h_{A_2}/h_{A_1}$ and $h_{-}/h_{+}$.
In order to estimate the probable size of those missing corrections, we have made the following replacements in eqs. \ref{Riws} and \ref{fpbyf0}:
\begin{align}
\frac{h_{v}}{h_{A_1}} &\to \frac{h_{v}}{h_{A_1}} \Delta_v, \ \   \frac{h_{A_3}}{h_{A_1}} \to \frac{h_{A_3}}{h_{A_1}} \Delta_{31}, \ \  
 \frac{h_{A_2}}{h_{A_1}} \to \frac{h_{A_2}}{h_{A_1}} \Delta_{21},\ \ \ \  \frac{h_{-}}{h_{+}} \to \frac{h_{-}}{h_{+}} \Delta_{\mp}.
\end{align}
Though these $\Delta$s could be $w$ dependent in general, for values of $w$ very close to 1, these dependencies can be neglected. Thus we 
assume these $\Delta$s to be $w$ independent for simplicity.
Following the above-mentioned methods, we then fit these additional parameters along with the HQET sub-leading Isgur-Wise functions using all the available 
datasets as mentioned in table \ref{casescln}, with a goal to find out the size of these newly introduced parameters.
In our analysis, $\Delta$s are treated as normally distributed nuisance parameters with $\Delta = 1.0 \pm 0.2$, i.e. we have allowed these missing 
corrections to vary at most by 20\%.

Table \ref{tab:hqetcln} (3rd and 5th column) shows the fit results of the HQET parameters and the $\Delta$s with the corresponding 1$\sigma$ errors, 
for the scenarios listed in table \ref{casescln}. We note that $\Delta_{\mp}$ and $\Delta_{21}$ could be as large as 20\%, while $\Delta_v$ and $\Delta_{31}$ 
are $<$ 15\%. The set of parameter values thus obtained is used to estimate the best-fit values and uncertainties in $R_1(1)$ 
and $R_2(1)$. In figure \ref{fig:R1R2}, these fitted results for $R_{1(2)}(1)$ are compared with those obtained previously from the CLN fit and with the 
QCDSR predictions. Our CLN fit results of $R_1(1)$ and $R_2(1)$ have large uncertainties, and are marginally consistent with the QCDSR predictions, 
whereas those obtained using our fit results for the HQET parameters without $\Delta$s (i.e. cases 1 and 2 of \ref{tab:hqetcln}) have small errors 
(shown as solid black bars) and lie in between the CLN fit results and QCDSR predictions. In the same plot, the solid red bars 
represent the best fit values and the error bars of $R_1(1)$ and $R_2(1)$, which are obtained using the parameter values of the 
$\Delta_v$, $\Delta_{21}$, and $\Delta_{31}$ and the other HQET parameters as given in table \ref{tab:hqetcln}. With these sets of parameters, we can now fully 
reproduce the CLN fit results, and hence, will be marginally consistent with the QCDSR predictions. However, if we take $\Delta_v = 1 \pm 0.1$, 
$\Delta_{21} = 1 \pm 0.2$, and $\Delta_{31} = 1 \pm 0.2$ (conservative estimates\footnote{Here, we have used the maximum attainable values of the 
$\Delta$s, which are allowed by the data. From the fit, the allowed values of $\Delta_{31}$ is $<$ 15\%, however, for the conservative estimates 
we have allowed it to be as large as 20\%.}) then we can fully reproduce the CLN fit results, as well as the QCDSR predictions; the results are shown 
as dot-dashed red bars in figure \ref{fig:R1R2}. Using the results given in table \ref{tab:hqetcln}, we estimate $R_0(1)$ for the above mentioned different 
cases. With these $R_0(1)$s and the CLN fit results given in table \ref{tab:resFcomb}, we obtain $R(D^*)$ which are presented in table \ref{tab:clnrdst}.
The errors in the estimated $R(D^*)$ have increased from 1.16\% to 2.32\% due to introduction of an additional error of about 20\% in the HQET form factor ratios.

\section{BGL parametrization  }\label{bgl}
\subsection{Formalism and Results for $|V_{cb}|$}

\begin{table}[!hbt]
 \begin{center}
 \def\arraystretch{1.4}
  \begin{tabular}{|c|ccc|}
     \hline
     \multicolumn{4}{|c|}{Data+Lattice (HPQCD \& MILC) }  \\
     \hline
       Parameters/ & Best Fit & $\pm$ Err& Err. from \\
       Observables & Values   &          & $\Delta \chi^2 = \pm 1$ \\ 
      \hline
 $|V_{cb}|\times 10^{3}$ & 41.04 & 1.13 &  \errvec{+1.12 \\ -1.13} \\
 \hline
 $a^{f_+}_0$ & 0.0141 & 0.0001 &  (0.0001) \\
 $a^{f_+}_1$ & -0.0318 & 0.0028 &  (0.0028)  \\
 $a^{f_+}_2$ & -0.0819 & 0.0199  & (0.0199) \\
 $a^{f_0}_1$ & -0.1961 & 0.0136 & (0.0136) \\
 $a^{f_0}_2$ & -0.2274 & 0.0942  & (0.0942) \\
 \hline
 $\chi^2_{min} $ & \multicolumn{3}{c|}{33.37}  \\
 dof & \multicolumn{3}{c|}{46} \\
 $p$-value & \multicolumn{3}{c|}{91.77\%}  \\
 \hline
  R(D) & \multicolumn{3}{c|}{ $0.302$ $\pm 0.003$ }\\ 
 \hline
 \end{tabular}
 \end{center}
 \caption{The fit results obtained from the analysis of the decay \bdlnu with the BGL parameterization of the form factors for $N=2$.}
 \label{tab:bdlnuBGL}
\end{table}

\begin{table}[!t]
 \begin{center}
 \def\arraystretch{1.3}
  \begin{tabular}{|c|cc|cc|}
     \hline
     &\multicolumn{2}{c|}{Data+Lattice} & \multicolumn{2}{c|}{Data+Lattice+LCSR} \\
     \cline{2-5}
       Parameters & Best Fit & Err. from & Best Fit & Err. from \\
      & Values & $\Delta \chi^2 = 1$ & Values & $\Delta \chi^2 = 1$ \\
      \hline
 $|V_{cb}|\times 10^{3}$ & 41.7 & \errvec{+ 2.0 \\ -2.1} & 40.6 &  (1.7) \\
 \hline
 $a^f_0$ & 0.0109 & (0.0002) & 0.0109 & (0.0002) \\
 $a^f_1$ & -0.0459 & \errvec{+0.0527 \\ -0.0429} & -0.0518 & \errvec{+0.0267 \\ -0.0131} \\
 $a^f_2$ & 0.1513 & \errvec{0.8457 \\ -1.1508} & 0.9942 & \errvec{+0.0047 \\ -0.5019} \\
 \hline
 $a^{\mathcal{F}_1}_1$ & -0.0092 & \errvec{+0.0054 \\ -0.0050} & -0.0070 & \errvec{+0.0048 \\ -0.0046} \\
 $a^{\mathcal{F}_1}_2$ & 0.1150 & \errvec{+0.0877 \\ -0.0921} & 0.0932 & \errvec{+0.0850 \\ -0.0883} \\
 \hline
 $a^g_0$ & 0.0111 & \errvec{+0.0104 \\ -0.0075} & 0.0257 & \errvec{+0.0054 \\ -0.0034} \\
 $a^g_1$ & 0.5786 & \errvec{+0.3351 \\ -0.4007} & 0.0836 & \errvec{+0.0753 \\ -0.2157} \\
 $a^g_2$ & 0.8155 & \errvec{+0.1683 \\ -1.7701} & -0.9962 & \errvec{+1.9958 \\ -0.0036} \\
      \hline
 $\chi^2_{min} $ & \multicolumn{2}{c|}{27.81} & \multicolumn{2}{c|}{30.93} \\
 dof & \multicolumn{2}{c|}{32} & \multicolumn{2}{c|}{35} \\
 $p$-value & \multicolumn{2}{c|}{67.87\%} & \multicolumn{2}{c|}{66.51\%} \\
 \hline
 \end{tabular}
 \end{center}
 \caption{Fit results with BGL parameterization of the form factors ($N=2$) in \bdstlnu .}
 \label{tab:bdstlnuBGL}
\end{table}

The BGL parameterization of the form factors rely on a Taylor series expansion about $z=0$. The key ingredient in this approach is the 
transformation that maps the complex $q^2$ \footnote{In our case, $q^2 = (p_{\ell} + p_\nu)^2$} plane onto the unit disc $|z| \le 1$. The most general form of 
the expansion of the form factors is given as \cite{BGL}
\begin{equation}
	F_i(z) = \frac{1}{P_{i}(z )\phi_{i}(z )}\sum_{n=0}^N a_n^{{\cal F}_i}\ z^n,
	\label{bglform}
\end{equation}
where
\begin{equation}
	z=\frac{\sqrt{w +1}-\sqrt{2}}{\sqrt{w +1}+\sqrt{2}}. 
\end{equation}
Here, $F_i(z)$ include $f_+(z)$, $f_0(z)$, associated with the decays $B\to D$, and those associated with $B\to D^*$ are given by $F_1(z)$, $f(z)$, $g(z)$
and $F_2(z)$ respectively. The coefficients $a_n^{{\cal F}_i}$ follow weak as well as strong unitarity constraints \cite{BGL}.
However, along the lines of ref.s \cite{Bigi:2016mdz,Bigi:2017,BGL}, we have used the 
weak unitarity constraints for the coefficients $a_n^{f_+}$, $a_n^{f_0}$, $a_n^{{\cal F}_1}$, $a_n^{f}$, $a_n^{ g}$ and $a_n^{{\cal F}_2}$, and considered 
the form factors with $N=2$ in our analysis. The weighting functions $\phi_i(z)$ contain the Jacobian of the variable transformation and the physics of the 
perturbative QCD (PQCD). It is also analytic on the unit disc. The mathematical forms of these $\phi_i$'s, corresponding to 
various spin states, can be seen from \cite{BGL}. Another important ingredient in this form of parameterization is the Blaschke factor, which is defined as
\begin{equation}
	P(z) = \prod_p \frac{z-z_p}{1 - z z_p},  
\end{equation}
where 
\begin{equation}
	z_p =\frac{\sqrt{t_+ - m_{p}^2} - \sqrt{t_+ - t_0}}{\sqrt{t_+ - m_{p}^2} + \sqrt{t_+ - t_0}},
\end{equation}
with 
\begin{equation}
	t_+=(m_B + m_{D^{(*)}})^2 , t_-=(m_B - m_{D^{(*)}})^2 , t_0 = t_-
\end{equation}
Here, $ z = z_p$ represents the location of a pole i.e $B_c$ narrow resonance. The $P(z)$ is analytic on the unit disc for $|z_p| \le 1$. In general, the form 
factors $F_i(q^2)$ have poles, and the Blaschke factor is useful to eliminate those poles of $F_i$'s at $z =z_p$, such that $P_i F_i$ is analytic on the unit 
disc $|z| \le 1$. 

In our analysis, the various inputs relevant to the BGL parameterization of the form factors associated with the decays \bdlnu and \bdstlnu  
($\ell = \mu$ and e) are taken from the references \cite{Bigi:2016mdz,Bigi:2017,BGL}. The fit results are shown in Tables \ref{tab:bdlnuBGL} and 
\ref{tab:bdstlnuBGL} respectively. The results of the combined analysis of data in $B\to D^{(*)} \ell\nu_{\ell} $ are shown in Table \ref{tab:combBGL}. 
In the combined analysis, we use the following weak unitarity constraints:
\begin{align}
        (a_0^{g})^2 + (a_1^{g})^2 + (a_2^{g})^2 + (a_0^{f_+})^2 + (a_1^{f_+})^2 + (a_2^{f_+})^2 < 1, \nn \\
        (a_0^{{\cal F}_1})^2 + (a_1^{{\cal F}_1})^2 + (a_2^{{\cal F}_1})^2 + (a_0^{f})^2 + (a_1^{f})^2 + (a_2^{f})^2 < 1, \nn \\
        (a_0^{f_0})^2 + (a_1^{f_0})^2 + (a_2^{f_0})^2 < 1.
\end{align}

We note that the uncertainties of the extracted $|V_{cb}|$ have reduced to $\approx$ 2\%. This is 
the most precise estimate obtained so far from a combined analysis. In addition, we observe that the central values of the $|V_{cb}|$ obtained from BGL analysis is 
increased by approximately 3.5\% for combined fit without LCSR and 3\% for combined fit with LCSR than those obtained using the CLN parameterizations for 
the form factors. 

\begin{table}[!t]
 \begin{center}
 \def\arraystretch{1.3}
  \begin{tabular}{|c|cc|cc|}
     \hline
     & \multicolumn{2}{c|}{Data+Lattice} & \multicolumn{2}{c|}{Data+Lattice+LCSR} \\
     \cline{2-5}
       Parameters & Best Fit & Err. from & Best Fit & Err. from \\
      & Values & $\Delta \chi^2 = 1$ & Values & $\Delta \chi^2 = 1$ \\
      \hline
 $|V_{cb}|\times 10^{3}$ & 41.2 & (1.0) & 40.9 & (0.9) \\
 \hline
 $a^f_0$ & 0.0109 & (0.0002) & 0.0109 & (0.0001) \\
 $a^f_1$ & -0.0366 & \errvec{+0.0409 \\ -0.0422} & -0.0534 & \errvec{+0.0194 \\ -0.0112} \\
 $a^f_2$ & -0.0340 & \errvec{+1.0312 \\ -0.9652} & 0.9936 & \errvec{+0.0049 \\ -0.4022} \\
 \hline
 $a^{\mathcal{F}_1}_1$ & -0.0084 & \errvec{+0.0045 \\ -0.0044} & -0.0074 & \errvec{+0.0043 \\ -0.0042} \\
 $a^{\mathcal{F}_1}_2$ & 0.1054 & \errvec{+0.0846 \\ -0.0855} & 0.0983 & \errvec{+0.0821 \\ -0.0830} \\
 \hline
 $a^g_0$ & 0.0112 & \errvec{+0.0108 \\ -0.0075} & 0.0256 & \errvec{+0.0052 \\ -0.0033} \\
 $a^g_1$ & 0.5882 & \errvec{+0.3320 \\ -0.4233} & 0.0800 & \errvec{+0.0722 \\ -0.2131} \\
 $a^g_2$ & 0.8038 & \errvec{+0.1783 \\ -1.7582} & -0.9925 & \errvec{+1.9887 \\ -0.0038} \\
 \hline
 $a^{f_+}_0$ & 0.0141 & (0.0001) & 0.0141 & (0.0001) \\
 $a^{f_+}_1$ & -0.0320 & (0.0027) & -0.0317 & (0.0027) \\
 $a^{f_+}_2$ & -0.0816 & (0.0199) & -0.0822 & (0.0198) \\
 \hline
 $a^{f_0}_1$ & -0.1967 & (0.0134) & -0.1956 & (0.0134) \\
 $a^{f_0}_2$ & -0.2291 & (0.0941) & -0.2259 & (0.0940) \\
      \hline
 $\chi^2_{min} $ & \multicolumn{2}{c|}{61.26} & \multicolumn{2}{c|}{64.35} \\
 dof & \multicolumn{2}{c|}{79} & \multicolumn{2}{c|}{82} \\
 $p$-value & \multicolumn{2}{c|}{93.04\%} & \multicolumn{2}{c|}{88.35\%} \\
 \hline
 \end{tabular}
 \end{center}
 \caption{The fit results obtained from the combined analysis of the available data in the decays \bdlnu and \bdstlnu with the BGL parameterization of the 
 form factors with $N=2$.}
 \label{tab:combBGL}
\end{table}

\subsection{Predictions for $R(D^{(*)})$}
As mentioned earlier, the decay $B\to D^* \tau\nu_{\tau}$ is sensitive to an additional form factor \cite{BGL} 
\begin{equation}
	F_2(z) =\frac{1}{P_{2}(z)\phi_{2}(z)}\sum_{n=0}^N a_n^{{\cal F}_2}\ z^n
	\label{f2z}
\end{equation}
with 
\begin{equation}
	P_{2}(z)=\prod_{p=1}^3 \frac{z - z_p }{1 - z z_p },
\end{equation}
and 
\begin{equation}
	\phi_{2}(z)=\sqrt{\frac{2 n_I}{\pi \tilde{\chi}_{0^-}}}
	\frac{2^3 (r)^2 (1+z)^2 (1-z)^{-\frac{1}{2}}}{[(1+r)(1-z)
		+2\sqrt{r}(1+z)]^4}.
\end{equation}
Here, $n_I$ represents the number of light valence quarks or the effective iso-spin factor, as given in \cite{Bigi:2016mdz}. We use $n_I=2.6$. The functions 
$\tilde{\chi}_{0^-}$ are defined as \cite{CLN}
\begin{equation}
	\tilde{\chi}_{0^-}(0)=\chi_{0^-}(0)-\sum_{n=1}\frac{f^2_n(B_c)}{M^2_n(B_c)},
\end{equation}
where $\chi_{0^-}(0)$ are the perturbatively calculable functions and associated with the once-subtracted QCD dispersion relation ( for details see \cite{BGL}). 
The second term represents the contribution to the dispersion relations from the $0^-$ $B_c$-type resonances below the $B D^*$ type pair production. 
The decay widths and masses of those resonances are given by $f_n$ and $M_n$ respectively, and their respective values are given in Table 
\ref{tab:resonances}. 

\begin{table}[!hbt]
	\begin{center} 
		\begin{tabular}{|cccc|}
			\hline
			Form- & Resonance & Mass, & Decay constant, \\
			factor    & type  & $M_n$ in GeV & $f_n$ in GeV \\
			\hline
			& & 6.275 & 0.427\\
			$F_2$ & $0^-$ & 6.842 & \\
			& & 7.250 & \\
			\hline
		\end{tabular}
	\end{center}
	\caption{The decay widths and the masses of the $B_c$ resonances.}
	\label{tab:resonances}
\end{table}

Expressions of $\chi_{0^-}(0)$ include the corrections at order $\alpha_s$. We obtain $\chi_{0^-} = (1.807 \pm 0.009) \times 10^{-2}$ using the inputs given 
in eq. \ref{thinputs}. However, incorporating the corrections of order $\alpha^2_s$, one will obtain $\chi_{0^-} =  1.942 \times 10^{-2}$ \cite{Bigi:2017jbd}. 

In eq. \ref{f2z}, the unknowns are the various coefficients $a^{{\cal F}_2}_n$. For $N=2$ these are, respectively, 
$a^{{\cal F}_2}_0$, $a^{{\cal F}_2}_1$ and $a^{{\cal F}_2}_2$. Hence, in order to predict $R(D^*)$ one needs to extract these coefficients, and HQET relations 
between the form factors are very useful in this regard. In order to extract these coefficients, we use the following equations 
\begin{equation}
	F_2(w) = \left(\frac{F_2(w)}{F_i(w)}\right)_{\scaleto{HQET}{3pt}} F_i(w), \ \ i\ne 2. 
	\label{calcoeff}
\end{equation}
Here, $F_i(w)$'s can be anyone of $f_+(w)$, $f_0(w)$, $F_1(w)$ and $f(w)$.   
As mentioned earlier, the HQET form factors at order $\alpha_s$ and $\Lambda_{QCD}/m_{b,c}$ are given in terms of the five HQET parameters  
(for details see \cite{Bernlochner:2017}). For the known values of these parameters, the r.h.s of eq. \ref{calcoeff} are different numbers for different 
values of $w$ ($\ge 1$), since $F_i(w)$ are known, either from our fits or from the lattice. Hence, we can create synthetic data points for $F_2(w)$ for different
values of $w$.

\begin{table}[t]
	\begin{center}
		\begin{tabular}{|c|c|}
			\hline
			Cases  & Inputs for the fits \\
			\hline
			case-3  & $F_1(w)/f(w)$ for $w$=1.03, 1.06, 1.09 and\\
			&  $f_+(w)/f_0(w)$ for  $w$=1, 1.03, 1.06, 1.09 \\
                               & from BGL fit results (table \ref{tab:combBGL}) \\
                               \hline
			case-4 & case-3 with $R_1(w_{max})$ and $R_2(w_{max})$  \\
			&  from LCSR \\
			\hline
			case-5  & $f_+(w)/f_0(w)$  for $w$=1, 1.08, 1.16 (MILC) \\
                                &   and $w$=1.03, 1.06, 1.09, 1.12 (HPQCD) \\
                                \hline
			case-6 & case-5 with $R_1(w_{max})$ and $R_2(w_{max})$  \\
			&  from LCSR \\
			\hline        
		\end{tabular}
	\end{center}
	\caption{Different cases for the fit of sub-leading Isgur-Wise functions.}
	\label{cases}
\end{table}

\begin{table}[!t] 
 \begin{center}
  \def\arraystretch{1.5}
  \begin{tabular}{|c | c | c |c | c | c | c | c | c| }
     \hline
     Para- & case-3    & case-3 & case-4 & case-4 & case-5 &  case-5   &  case-6 & case-6    \\
     meters &    & with $\Delta$s&     & with $\Delta$s&     & with $\Delta$s&  & with $\Delta$s  \\
     \hline
     $\eta(1)$         & 0.39(3) & 0.39(5)  & 0.38(3)  & 0.40(5) & 0.39(4) & 0.40(5)  & 0.40(3) & 0.40(5)     \\
$\eta^{\prime}(1)$     & 0.10(7) & 0.12(5) & 0.08(7)  & 0.14(7) & 0.01(12) & 0.004(101) & -0.02(10) & 0.003(101)   \\
$\chi_2(1)$            & -0.07(6) & -0.05(6)& -0.11(5) & -0.08(6)  & -0.06(6) & -0.06(6)   &-0.06(6) & -0.06(6)   \\
${\chi_2}^{\prime}(1)$ & 0.007(60)& -0.02(4) & 0.006(59) & -0.004(30) & -0.003(60) & -0.003(60)  &-0.002(59)& -0.003(60)  \\
${\chi_3}^{\prime}(1)$ & 0.06(5) & 0.06(4) & 0.06(5)  & 0.04(4) & 0.04(6) & 0.04(6)  & 0.05(6)&  0.04(6) \\
\hline
$\Delta_v$    & -  &    -      &  -  & 1.06(3) & - &  -     & - & 1.06(3)     \\  
$\Delta_{\mp}$ & -  & 0.98(20)  & -  & 1.00(20)& - & 1.00(20)& - &  1.00(20) \\  
$\Delta_{21}$ & -  & 1.05(20)  & -   & 1.02(20)& -& -       & - & 1.00(20)) \\  
$\Delta_{31}$ & -  & 1.03(10)  & -   & 1.07(7) & -& -       & - & 1.01(13)       \\ 
      \hline    
$\chi^2_{min} $ & 1.71 & 1.73 & 7.63 & 1.88   & 3.84  & 3.26   & 7.05 & 3.36 \\
 dof           & 5      & 5    & 7     & 7      & 5    &  5       &  7   & 7   \\
 $p$-value & 88.77\% & 88.54\%  & 36.62\%  & 96.60\% & 57.25\%  & 66.02\% & 42.36\% & 84.97\% \\ 
 \hline
 \end{tabular}
 \end{center}
 \caption{The fit results for the subleading Isgur-Wise parameters with $\chi_2(1)$, $\chi_2^{\prime}(1)$ and $\chi_3^{\prime}(1)$ varied within their QCDSR 
 $3~ \sigma$ range.}
 \label{tab:hqet}
\end{table}

\begin{table}[t] 
	\begin{center}
		\def\arraystretch{1.5}
		\begin{tabular}{|c | c | c |c | c | c | c | c | c| }
			\hline
			Para- & case-3    & case-3 & case-4 & case-4 & case-5 &  case-5   &  case-6 & case-6    \\
			meters &    & with $\Delta$s&     & with $\Delta$s&     & with $\Delta$s&  & with $\Delta$s  \\
			\hline
			$\eta(1)$         & 0.39(3)  & 0.40(4)    & 0.37(3)  & 0.40(4)    & 0.39(3)    & 0.40(5)    & 0.39(3)   & 0.40(5)     \\
			$\eta^{\prime}(1)$     & 0.13(4)  & 0.13(4)    & 0.11(4)  & 0.14(4)    & 0.005(105) & 0.003(92) & -0.02(10) & 0.002(92)   \\
			$\chi_2(1)$            & -0.06(2) & -0.06(2)   & -0.07(2) & -0.06(2)   & -0.06(2)   & -0.06(2)   &-0.06(2)   & -0.06(2)   \\
			${\chi_2}^{\prime}(1)$ & 0.001(20)& -0.003(14) & 0.001(20)& -0.004(14) & -0.000(20) & -0.000(20) &-0.000(20) & -0.000(20)  \\
			${\chi_3}^{\prime}(1)$ & 0.04(2)  & 0.04(2)    & 0.04(2)  & 0.04(2)    & 0.04(2)    & 0.04(2)    & 0.04(2)   &  0.04(2) \\
			\hline
			$\Delta_v$    & -  &    -      &  -  & 1.06(3) & - &  -     & - & 1.06(3)     \\  
			$\Delta_{\mp}$ & -  & 1.00(19)  & -  & 0.98(19)& - & 1.00(20)& - &  1.00(20) \\  
			$\Delta_{21}$ & -  & 1.00(20)  & -   & 0.98(20)& -& -       & - & 1.00(20) \\  
			$\Delta_{31}$ & -  & 1.03(10)  & -   & 1.08(6) & -& -       & - & 1.01(13)       \\ 
			\hline    
			$\chi^2_{min} $ & 1.92 & 1.68 & 8.58 & 1.80   & 3.84  & 3.26   & 7.36 & 3.36 \\
			dof           & 5      & 5    & 7     & 7      & 5    &  5       &  7   & 7   \\
			$p$-value & 85.97\% & 89.07\%  & 28.39\%  & 97.02\% & 57.19\%  & 65.96\% & 39.26\% & 84.94\% \\ 
			\hline
		\end{tabular}
	\end{center}
	\caption{The fit results for the subleading Isgur-Wise parameters with $\chi_2(1)$, $\chi_2^{\prime}(1)$ and $\chi_3^{\prime}(1)$ varied within their QCDSR $1~ \sigma$ range.}
	\label{tab:hqet1sig}
\end{table}

\begin{figure*}[htbp]
	\centering
	\includegraphics[scale=0.4]{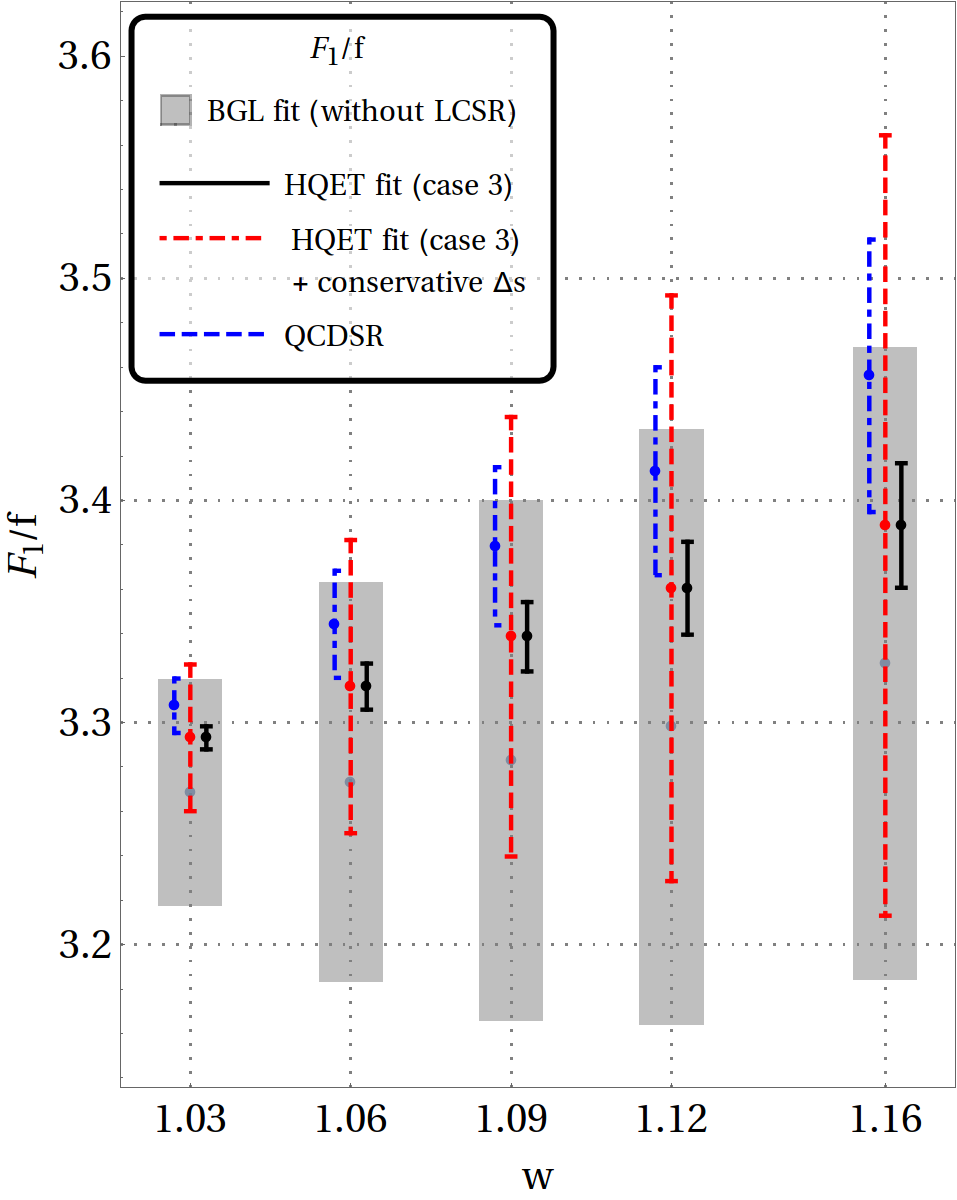}~~~~~~~~~~~~
	\includegraphics[scale=0.4]{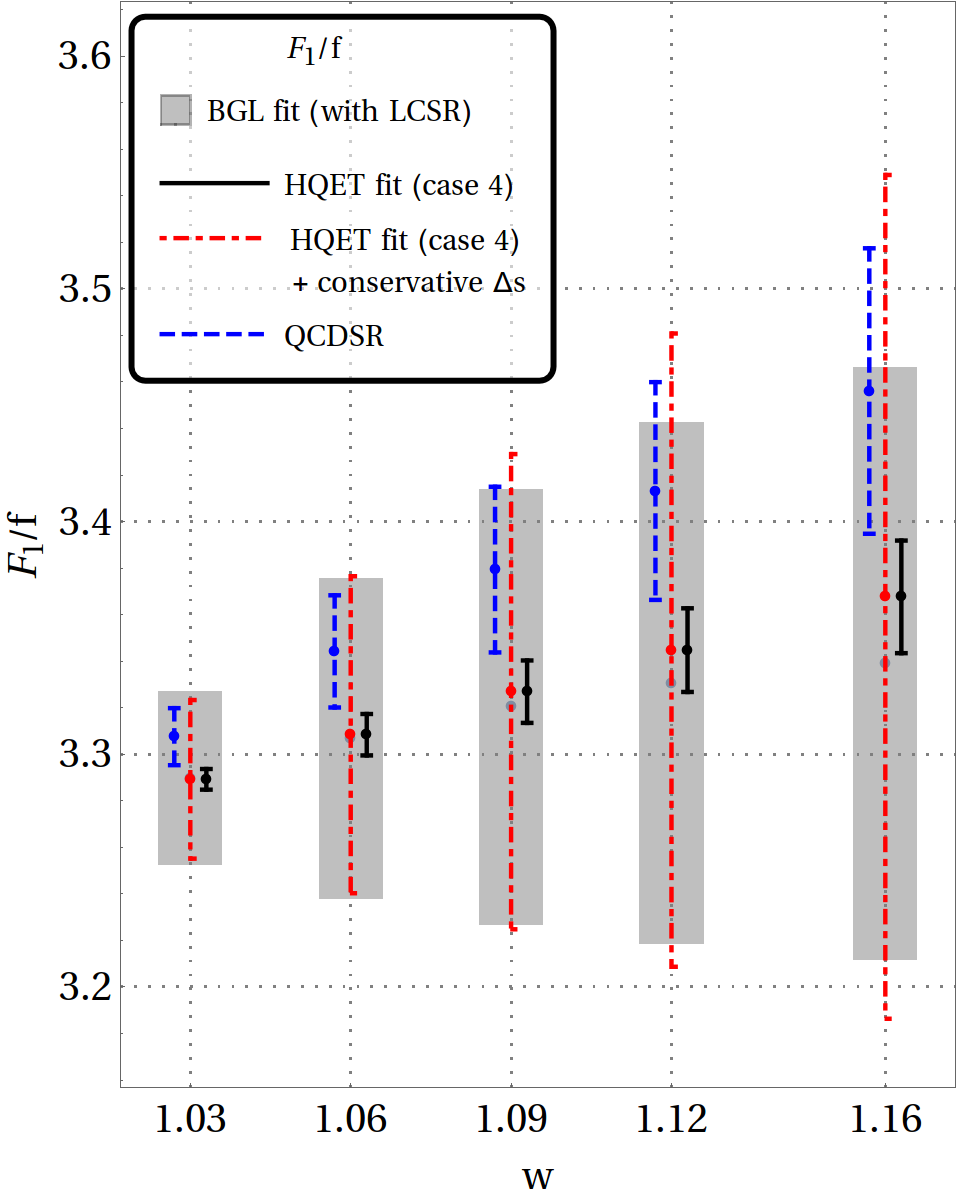} \\
	\includegraphics[scale=0.4]{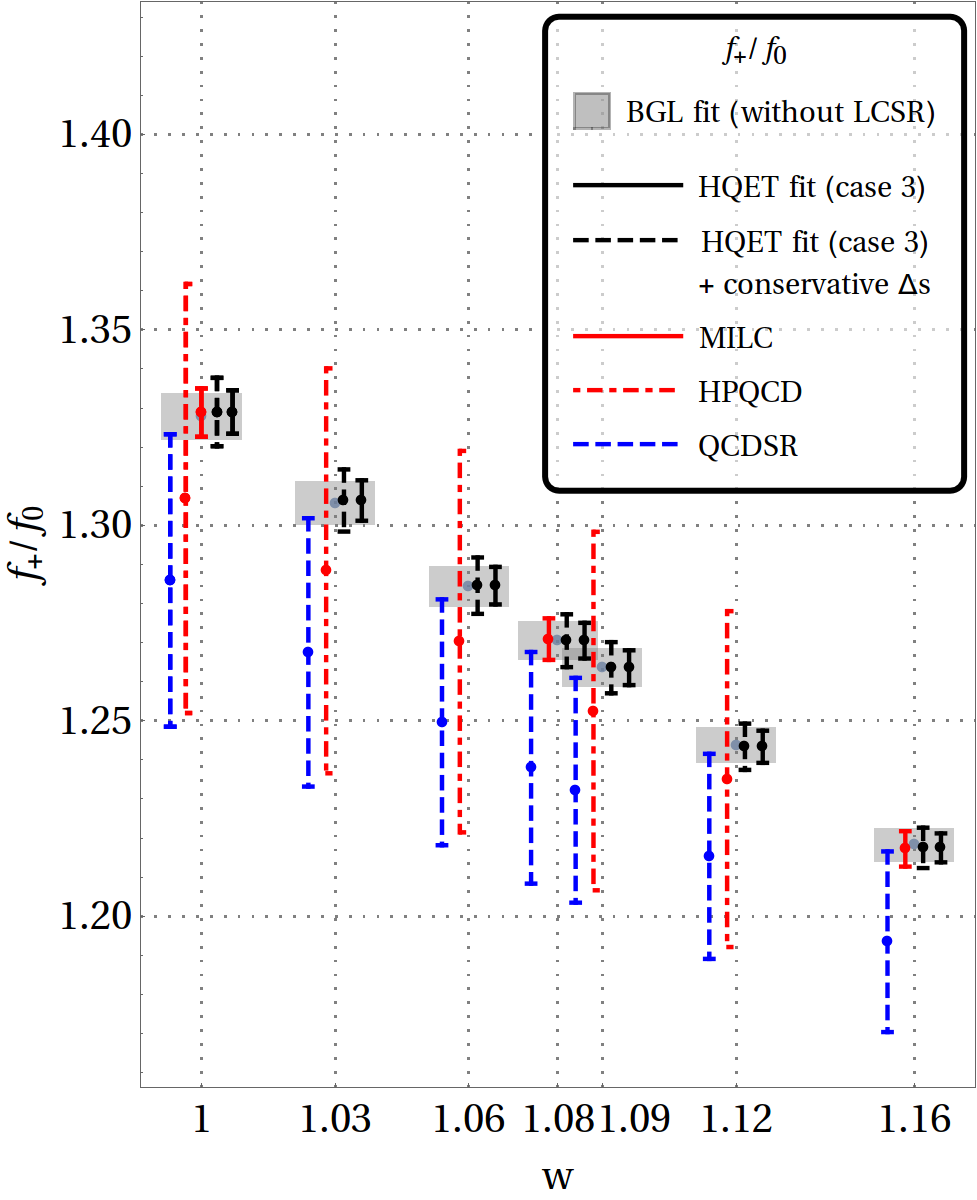}~~~~~~~~~~~~
	\includegraphics[scale=0.4]{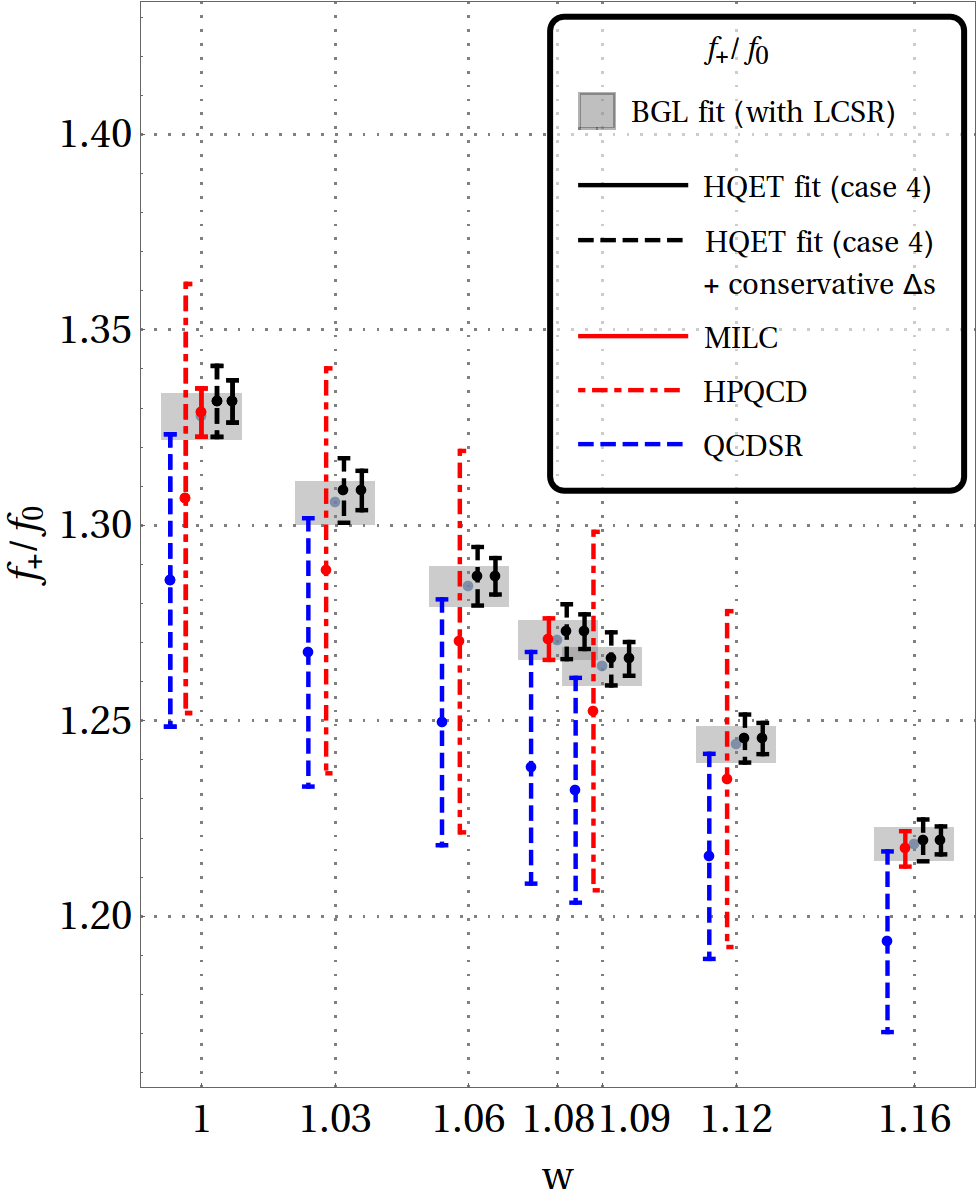} \\
	\caption{Upper half panel: Comparisons between different fit results of $F_1(w)/f(w)$ (1$\sigma$ bars) with (right) and without (left) the inputs from LCSR. 
		Lower half panel: The ratio $f_{+}/f_0$ obtain from different fits, and from lattice. The fit results are also compared with the QCDSR predictions.}
	\label{fig:F1byf}
\end{figure*}

In the HQET, the form factor ratios $F_1(w)/f(w)$ and $F_2(w)/f(w)$ are given by
\begin{align}\label{F1byf}
 \frac{F_1(w)}{f(w)} &= m_B (w-1)\left(\frac{w - r_{D^*}}{w-1} - \frac{h_{A_2}}{h_{A_1}} r_{D^*} - \frac{h_{A_3}}{h_{A_1}} \right),  \nn \\
 \frac{F_2(w)}{f(w)} &= \frac{1}{m_B r_{D^*}}\left( 1 - \frac{h_{A_2}}{h_{A_1}} \frac{1- r_{D^*} w}{1+w} - \frac{h_{A_3}}{h_{A_1}} \frac{w - r_{D^*}}{1+w} \right)\,. \nn \\
 \end{align}
We can also define
\begin{equation}\label{F2byF1} 
\frac{F_2(w)}{F_1(w)} = \frac{F_2(w)}{f(w)} \frac{f(w)}{F_1(w)},
\end{equation}
such that all these form factor ratios are sensitive to $h_{A_2}/h_{A_1}$ and  $h_{A_3}/h_{A_1}$. 
The other form factor ratios, which can also be used in the extractions of $R(D^*)$ are given by
\begin{align}\label{F2byfp}
\frac{F_2(w)}{f_+(w)} &= 2\frac{\left(1 +w - \frac{h_{A_2}}{h_{A_1}} (1  - r_{D^*} w) + 
\frac{h_{A_3}}{h_{A_1}} (r_{D^*} -  w) \right)}
 {\frac{\sqrt{r_{D^*}}}{\sqrt{r_{D}}} \frac{h_+}{h_{A_1}}(\frac{h_-}{h_+} (r_D-1) + (1 + r_D))}, \nn \\ 
\frac{F_2(w)}{f_0(w)} &=  \frac{\left(1 - \frac{h_{A_2}}{h_{A_1}} (1  - r_{D^*} w) + 
\frac{h_{A_3}}{h_{A_1}} (r_{D^*} -  w) \right)}
 {\frac{\sqrt{r_{D^*}}\sqrt{r_{D}}}{r_D +1} \frac{h_+}{h_{A_1}}(\frac{h_-}{h_+} \frac{(r_D+1)(w-1)}{(r_D-1)(w+1)} + 1 )}.
 \end{align}
 Apart from $h_{A_2}/h_{A_1}$ and  $h_{A_3}/h_{A_1}$, these ratios are sensitive to $h_-/h_+$ and $h_+/h_{A_1}$. 
  
Using the fit results given in table \ref{tab:combBGL}, we generate the synthetic data points for the ratios
$F_1(w)/f(w)$ and $f_+(w)/f_0(w)$, for different values of $w (\ge 1)$. We first fit the sub-leading Isgur-Wise functions using those synthetic data points. 
In the analysis, the different benchmark cases are defined in table \ref{cases} and the respective fit results are shown in table \ref{tab:hqet}.
The lattice inputs are playing the major role in all the fits, and all the fits are good with physically plausible values for the HQET parameters. 
We note here that in general, the ratios of the form factors are more sensitive to $\eta(1)$ than the other HQET parameters. 
The values of the HQET parameters predicted in QCDSR have large errors ($\gsim$ 30\%). For all of our intents and purposes, while taking the range of these
parameters seriously, we do not regard their central values with similar import. Therefore, we had initially tried to fit all the HQET parameters from the data 
and lattice. In general, the fits were not good, and on top of that the error values of the parameters $\chi_2(1)$, $\chi_2^{\prime}(1)$ and $\chi_3^{\prime}(1)$ were 
very large, in some cases, the errors were almost eight times the corresponding best fit values. Also, the best fit values of these parameters were almost 
$\gsim$ 3$\sigma$ away than the respective QCDSR predictions. In all those fits, the parameter $\eta(1)$ had small errors. However, $\eta^{\prime}(1)$ had large errors 
but it was in accordance with the QCDSR prediction. 

Thus, in order to get a good fit of the data with reasonable and conservative uncertainties in the extracted parameters, we have considered
$\chi_2(1)$, $\chi_2^{\prime}(1)$ and $\chi_3^{\prime}(1)$ as nuisance parameters, but varied 
them as gaussians in a broader range of three times the uncertainties associated with QCDSR predictions 
\cite{Bernlochner:2017,Neubert:1992f,Neubert:1992s}, which means our inputs for the fits are 
\begin{equation}
 \chi_2(1) = -0.06 \pm 0.06, \ \ \chi_2^{\prime}(1) = 0 \pm 0.06, \ \  \chi_3^{\prime}(1) = 0.04 \pm 0.06. 
\end{equation}
The corresponding results are given in table \ref{tab:hqet}. For completeness, we have also provided the set of results
with the parameters varied as a nuisance gaussian over their predicted QCDSR range, in table \ref{tab:hqet1sig}. The relaxed ranges of 
$\chi_2(1)$, $\chi_2^{\prime}(1)$ and $\chi_3^{\prime}(1)$ of the former fit normally produces larger uncertainties on the fitted parameters (table \ref{tab:hqet}). 
This, while having virtually no effect on the $R(D^*)$ calculation (this fact can be checked by comparing tables \ref{tab:rdstfinal} and \ref{tab:rdstfinal1sig} ), 
increases the quality of the fits somewhat and we have used them in our analysis. In the upper half panel of the figure \ref{fig:F1byf}, various fit results of 
$F_1/f$ (1$\sigma$ bars) as well as the predictions from QCDSR for different 
values of $w$ are compared. The BGL fit results (table \ref{tab:combBGL}) have large uncertainties, while uncertainties obtained using the 
fitted HQET parameters (cases 3 and 4 from table \ref{tab:hqet}) are very small. As can be understood from the lower-half panel of figure \ref{fig:F1byf}, 
the tight constraints on the HQET parameters are mainly coming from the lattice results, in particular from the MILC collaboration, on $f_{+}(w)/f_0(w)$. Apart from 
the high $w$ values, the BGL fit results of $F_1/f$ 
are fully consistent with that of QCDSR predictions. However, the same ratios obtained using the HQET fit parameters are marginally consistent with the QCDSR predictions.

We define a few more cases in addition to those given in table \ref{cases} where we have introduced additional parameters $\Delta_{\mp}$, $\Delta_{31}$, 
$\Delta_{21}$, and $\Delta_{v}$ as before. Note that we have not used $g(w)$ in the extraction of HQET parameters, as the uncertainties in the fitted values of 
$g(w)$ are large compared to other form factors. However, the parameter $\Delta_v$ will appear in the calculation while using $R_1(w_{max})$ as 
input. In order to fit these $\Delta$s along with the HQET parameters, we have to use synthetic data points on these form factors ratios. The fit 
results are shown in table \ref{tab:hqet}, and the allowed values of the $\Delta$s are same as those obtained previously in the analysis of the CLN fit results 
along with the lattice (table \ref{tab:hqetcln}). Upon incorporating these results in eqs. \ref{F1byf}, we get the estimates of the probable size of the additional errors in 
$F_1/f$, $F_2/f$ and $F_2/F_1$. For the ratios $F_2(w)/f_{+/0}(w)$, we need to know the probable size of the additional error in $h_+/h_{A_1}$, which can be 
obtained from a comparison of lattice result of $f_0(1)/f(1)$ with that obtained from the HQET fit results or QCDSR. We find it to be 
approximately 10\%, and assume it to be same for all other values of $w$. We propagate all these errors and estimate the overall size of the $\Delta$ in 
$F_2(w)/f_{+/0}(w)$. In order to be conservative in further analysis, we choose $\Delta_{31} = 1 \pm 0.2$ and $\Delta_{21} = 1 \pm 0.2$, 
and reproduce the ratios $F_1(w)/f(w)$ which are shown in the upper pannel of figure \ref{fig:F1byf} by the dot-dashed red bars. As expected, we can now fully 
reproduce the QCDSR results and most parts of the BGL fit results.

\begin{table}[htbp]
\tiny
 \begin{center}
  \def\arraystretch{1.4}
  \begin{tabular}{|c | c |c| c|  c | c|  c | c | c| }
     \hline
     & \multicolumn{2}{c|}{$F_i(w)$ in eq. \ref{calcoeff} \ :} & \multicolumn{2}{c|}{$F_i(w)$ in eq. \ref{calcoeff} \ :} 
			& \multicolumn{2}{c|}{$F_i(w)$ in eq. \ref{calcoeff}\ :}  & \multicolumn{2}{c|}{$F_i(w)$ in eq. \ref{calcoeff}\ :} \\
			& \multicolumn{2}{c|}{$F_1(w)$ \& $f(w)$} & \multicolumn{2}{c|}{$F_1(w)$ \& $f(w)$} & \multicolumn{2}{c|}{$f_+(w)$ \& $f_0(w)$}& 
			\multicolumn{2}{c|}{$f_+(w)$ \& $f_0(w)$} \\
     \cline{2-9}
 Parameters/ & case-3   & case-3  & case-4 & case-4 & case-5  &  case-5   &  case-6 & case-6 \\
 Observables    &     & with &     & with &     & with &  & with  \\
      &   & $\Delta$s  &   & $\Delta$s &   & $\Delta$s &  & $\Delta$s \\
     \hline
$a_0^{{\cal F}_2}$ & 0.053(1) & 0.053(4)& 0.053(1)&0.053(5)& 0.058(1)&0.058(8) & 0.058(1)  &0.058(8)      \\
$a_2^{{\cal F}_2}$ & 0.21(6)& 0.21(8)& -0.14(3)&-0.17(10)&-0.48(1)&-0.42(2)&-0.39(1)&-0.33(1)     \\
 \hline         
 $R(D)$  & 0.302(3)  & 0.302(3) &  0.302(3)  &  0.302(3)  &  0.302(3) & 0.302(3) &  0.302(3)  & 0.302(3)  \\
 \hline
 $\bm{R(D^*)}$ &  {\bf 0.255(5)} & {\bf 0.255(5)}   & {\bf 0.257(5)}   & {\bf 0.257(5)}  & {\bf 0.258(5)} & {\bf 0.258(7)}  & {\bf 0.260(5)}  &  {\bf 0.260(7)} \\
\hline
 Corr($R(D)$ & 0.12  & 0.11 & 0.12  & 0.10 & 0.14 & 0.10  & 0.13 & 0.09 \\
   -$R(D^*)$)      &   &  &   &  &  &  &  & \\
 \hline
 \end{tabular}
 \end{center}
 \caption{The predictions for $R(D^{(*)})$ using the fit results of the HQET parameters given in table \ref{tab:hqet}. $a_1^{{\cal F}_2}$ is fixed
 using eq. \ref{qcdcoeff}. The additional error ($\Delta$) in the ratio $h_+/h_{A_1}$ (eq. \ref{F2byfp}) are considered as $1 \pm 0.1$ (for detail, see the text). 
 Also, wherever applicable, the $\Delta_{31}$, $\Delta_{21}$ and $\Delta_{\mp}$, they all are taken as $1 \pm 0.2$.  
 The details of the choices of $F_i(w)$ in different cases can be seen from the text.}
 \label{tab:rdstfinal}
\end{table}

\begin{table}[htbp]
	\tiny
	\begin{center}
		\def\arraystretch{1.4}
		\begin{tabular}{|c | c |c| c|  c | c|  c | c | c| }
			\hline
			& \multicolumn{2}{c|}{$F_i(w)$ in eq. \ref{calcoeff} \ :} & \multicolumn{2}{c|}{$F_i(w)$ in eq. \ref{calcoeff} \ :} 
			& \multicolumn{2}{c|}{$F_i(w)$ in eq. \ref{calcoeff}\ :}  & \multicolumn{2}{c|}{$F_i(w)$ in eq. \ref{calcoeff}\ :} \\
			& \multicolumn{2}{c|}{$F_1(w)$ \& $f(w)$} & \multicolumn{2}{c|}{$F_1(w)$ \& $f(w)$} & \multicolumn{2}{c|}{$f_+(w)$ \& $f_0(w)$}& 
			\multicolumn{2}{c|}{$f_+(w)$ \& $f_0(w)$} \\
			\cline{2-9}
			Parameters/ & case-3   & case-3  & case-4 & case-4 & case-5  &  case-5   &  case-6 & case-6 \\
			Observables    &     & with &     & with &     & with &  & with  \\
			&   & $\Delta$s  &   & $\Delta$s &   & $\Delta$s &  & $\Delta$s \\
			\hline
			$a_0^{{\cal F}_2}$ & 0.053(1) & 0.053(4)& 0.053(1)&0.053(4)& 0.058(1)&0.058(8) & 0.058(1)  &0.058(8)      \\
			$a_2^{{\cal F}_2}$ & 0.19(3)& 0.02(9)& 0.18(6)& 0.08(19)&-0.72(1)&-0.46(15)&-0.73(12)&-0.46(15)     \\
			\hline         
			$R(D)$  & 0.302(3)  & 0.302(3) &  0.302(3)  &  0.302(3)  &  0.302(3) & 0.302(3) &  0.302(3)  & 0.302(3)  \\
			\hline
			$\bm{R(D^*)}$ &  {\bf 0.255(5)} & {\bf 0.255(5)}   & {\bf 0.257(5)}   & {\bf 0.257(5)}  & {\bf 0.258(5)} & {\bf 0.258(7)}  & {\bf 0.260(5)}  &  {\bf 0.260(7)} \\
			\hline
			Corr($R(D)$ & 0.12  & 0.11 & 0.12  & 0.10 & 0.14 & 0.10  & 0.13 & 0.09 \\
			-$R(D^*)$)      &   &  &   &  &  &  &  & \\
			\hline
		\end{tabular}
	\end{center}
	\caption{The predictions for $R(D^{(*)})$ using the fit results of the HQET parameters given in table \ref{tab:hqet1sig}. The rest of the assumptions are same 
	as that given in the caption of table \ref{tab:rdstfinal}.}
	\label{tab:rdstfinal1sig}
\end{table}

After the extraction of the HQET parameters, we use eq. \ref{calcoeff} to generate synthetic data points for $F_2(w)$ for different 
values of $w$ ($\ge 1$). In order to generate these synthetic data points one needs to find out $F_2(w)/F_i(w)$ for different values of $w$. As 
mentioned earlier, $F_i(w)$ could be anyone of $f(w)$, $F_1(w)$, and $f_{+/0}(w)$. The ratios $(F_2(w)/f(w))$ and $(F_2(w)/F_1(w))$ are less sensitive to the 
HQET parameters as compared to $(F_2(w)/f_{+/0}(w))$. Therefore, for case 3 (table \ref{cases}), we have replaced $F_i(w)$ in eq. \ref{calcoeff} by both $f(w)$ and $F_1(w)$ and created the 
above mentioned synthetic data points for $F_2(w)$. In case 4, the synthetic data points for $F_2(w)$ have been generated following the similar methods as are used
in case 3. For completeness, we have replaced $F_i(w)$ by $f_{+}(w)$ and $f_0(w)$ for creating the synthetic data points in case 5, and the similar normalizations 
are used in case 6. These synthetic data points for $F_2(w)$ are used in eq. \ref{f2z} to extract the coefficients $a_n^{{\cal F}_2}$ ( $n =0,1,2$ ).

Once the synthetic data points are generated, in all the cases, the coefficient $a_0^{{\cal F}_2}$ can be extracted directly by solving eq. \ref{f2z} for 
$w=1$ or $z=0$. Hence, the extracted values will be sensitive to $\eta(1)$ only. The extracted values of $a_0^{{\cal F}_2}$ which are shown in table 
\ref{tab:rdstfinal} and \ref{tab:rdstfinal1sig} are obtained by the use of the following synthetic data points (eq. \ref{calcoeff} for $w=1$) 
\footnote{We have checked that if we instead use $F_i(1) = F_1(1)$ (in cases 3 and 4), the value of 
$F_2(1)$ (and hence $a_0^{{\cal F}_2}$) remains exactly the same as that obatained with $F_i(1) = f(1)$. This is because both of $F_1(1)$ and $f(1)$  
are independent of HQET parameters and, in our BGL fits, we used the relation $F_1(1)=(m_B-m_{D^*})f(1)$. While using 
$F_i(1) = f_0(1)$ (in cases 5 and 6), the values of $F_2(1)$ and $a_0^{{\cal F}_2}$ had changed only slightly (unchanged at the precision we are quoting our results) 
with respect to the scenario $F_i(1) = f_+(1)$.}: 
\begin{equation}\label{F21}
  F_2(1)= 
  \begin{cases}
    \left(\frac{F_2(1)}{f(1)}\right)_{\scaleto{HQET}{3pt}} f(1) & \text{for cases 3 and 4},   \\
     \left(\frac{F_2(w)}{f_+(1)}\right)_{\scaleto{HQET}{3pt}} f_+(1) & \text{for cases 5 and 6} 
  \end{cases}
\end{equation}
In order to extract the other two coefficients, we have to use eq. \ref{f2z} for values of $w$ other than 1, and the unitarity constraint 
\begin{equation}
(a_0^{{\cal F}_2})^2 + (a_1^{{\cal F}_2})^2 + (a_2^{{\cal F}_2})^2 < 1 . 
\end{equation}
Naturally, the extracted values of $a_1^{{\cal F}_2}$ and $a_2^{{\cal F}_2}$ will be sensitive to the other HQET parameters along with $\eta(1)$. 
However, in order to reduce the impact of the HQET parameters on the final results, we use the QCD relation between the form factors:
\begin{equation}
 F_2(q^2=0) = \frac{2 F_1(q^2 = 0)}{m^2_B - m^2_{D^*}}.
 \label{qcdrelation}
\end{equation}
The coefficients of $F_1(z)$ are obtained from the BGL fits, and hence one of the coefficients in $F_2(z)$ can be written in terms of
the coefficients of $F_1(z)$ and the rest of the coefficients in $F_2(z)$ using the above relation, e.g., 
\begin{align}\label{qcdcoeff}
 a_1^{{\cal F}_2} &=  71.3906 a_0^f + 23.9092 a_1^{{\cal F}_1} + 1.34087 a_2^{{\cal F}_1}  - 17.8312 a_0^{{\cal F}_2}  - 0.0560815 a_2^{{\cal F}_2}.
 \end{align} 
Note that, $a_1^{{\cal F}_2}$ is highly sensitive to the extracted value of $a_0^{{\cal F}_2}$. However, for small values of $a_2^{{\cal F}_2}$ ( $\ll$ 1),  
 $a_1^{{\cal F}_2}$ has very little dependency on it. Also, $R(D^*)$ is relatively less sensitive to the coefficient $a_2^{{\cal F}_2}$ and its predictions do not 
 change depending on the changes in $a_2^{{\cal F}_2}$. Still, for completeness, we have extracted this coefficient from eq. \ref{f2z} by a fit using the synthetic data 
 points for $F_2(w)$ for $w$ = 1.03, 1.06, 1.09 and 1.12. As explained earlier, these $F_2(w)$ values are obtained using the following relations
 \footnote{We had also done the analysis using one normalization at a time. As for example, in both the cases 3 and 4, we had choosen 
 $F_i(w)$ to be either of $f(w)$ or $F_1(w)$. Similarly, for the cases 5 and 6, we had replaced $F_i(w)$ either by $f_+(w)$ or $f_0(w)$. In the specific cases, 
 we did not get any considerable changes in the predictions of $R(D^*)$ due to the different choices of the normalization of $F_2(w)$, also the predictions 
 were in complete agreement with those given in table \ref{tab:rdstfinal} and \ref{tab:rdstfinal1sig} which are obtained from the mixed normalizations. 
 However, for completeness, we have presented our results using mixed normalizations, it adds more inputs to the fits.}:  
 \begin{equation}
  F_2(w)= \left.
  \begin{cases}
    \left(\frac{F_2(w)}{f(w)}\right)_{\scaleto{HQET}{3pt}} f(w),   \\
    \left(\frac{F_2(w)}{F_1(w)}\right)_{\scaleto{HQET}{3pt}} F_1(w)  
  \end{cases}
  \right\}  \textmd{for both the cases 3 \& 4}
\end{equation}
and 
\begin{equation}
  F_2(w)= \left.
  \begin{cases}
    \left(\frac{F_2(w)}{f_+(w)}\right)_{\scaleto{HQET}{3pt}} f_+(w) ,  \\
    \left(\frac{F_2(w)}{f_0(w)}\right)_{\scaleto{HQET}{3pt}} f_0(w)  
  \end{cases}
  \right\}  \textmd{for both the cases 5 \& 6.}
\end{equation}
 The fitted values of $a_2^{{\cal F}_2}$ are shown in table \ref{tab:rdstfinal} and \ref{tab:rdstfinal1sig}.
 The coefficients  obtained in this way, and hence $R(D^*)$, will be mostly sensitive to $\eta(1)$. Therefore, the final results will be less dependent on 
 the HQET parameters.
 
We present our final results for $R(D^{(*)})$ in table \ref{tab:rdstfinal}. The prediction for $R(D)$ is consistent with the one obtained in an earlier analysis
\cite{Bigi:2016mdz}. Our important results are marked in bold.
Amongst these, the one obtained in case-4 (with $\Delta$) can be considered as our best result. The reasons are following:
\begin{itemize}
\item{In this case, the HQET parameters are fitted with all available inputs.}
\item{$R(D^*)$ has been extracted using the HQET relations $F_2(w)/f(w)$ and  $F_2(w)/F_1(w)$,
which are less sensitive to the HQET parameters (even $\eta(1)$) as compared to the other ratios, like $F_2(w)/f_{+/0}(w)$.}
\end{itemize}
We note that across all the cases, the overall uncertainties in predictions of $R(D^*)$ with the known corrections in HQET, are roughly 2\%. However, when we incorporate
the additional unknown corrections ($\Delta$s) conservatively in the HQET form factor ratios (from the fit), the uncertainties in cases 5 and 6 are increased to 3\%, 
while those in cases 3 and 4 remain the same. We have checked that in the cases 5 and 6 (with $\Delta$s), the overall uncertainties in the predictions of 
$R(D^*)$ are 4\% without using the QCD relation between the form factors, while those for the cases 3 and 4 (with $\Delta$s)
are roughly 3\%. Due to the QCD relation, the errors have reduced by 1\% in all these four cases with conservative $\Delta$s, 
which is due to a negative correlation between $a_0^{{\cal F}_2}$ and $a_1^{{\cal F}_2}$; for details see eq. \ref{qcdcoeff}.
The increase in errors for the cases 5 and 6 (with $\Delta$s) with respect to the cases 3 and 4 (with $\Delta$s) can be understood in the following way. 
The form factor ratios used in cases 5 and 6 have additional sources of errors compared to those used in the cases 3 and 4. As can be seen from 
eqs. \ref{F1byf}, \ref{F2byF1} and \ref{F2byfp}, in our analysis, the ratios $F_2(w)/f(w)$ and  $F_2(w)/F_1(w)$ are sensitive only to $\Delta_{21}$ and 
$\Delta_{31}$ while the ratios $F_2(w)/f_{+/0}(w)$ are sensitive to $\Delta_{21}$, $\Delta_{31}$, $\Delta_{\mp}$, and the additional unknown corrections 
associated with the ratio $h_{-}/h_{A_1}$. Our predictions for $R(D^*)$ are consistent with the one obtained in \cite{Bigi:2017jbd}.

\section{Summary}
In this article, we analyze the decay modes \bdlnu and \bdstlnu with the complete sets of available data on the angular (wherever applicable) as well $w$-bins.    
The CKM element $V_{cb}$ have been extracted from the analysis of the above mentioned decay modes independently, as well as from a combined analysis. We have done 
the analysis using the CLN and BGL parameterizations of the form factors. Our best results are $|V_{cb}| =39.77 \pm 0.89$ in the CLN parameterization of the form factors and 
that in the case of BGL is $|V_{cb}| = 40.90 \pm 0.94$. These are so far the most precise results obtained in the analysis of the exclusive decays. 
In the combined analysis of the data, our prediction for $R(D)$ in the CLN parameterization of the form factors is given by $R(D) = 0.304 \pm 0.003$, 
while using BGL parameterization, we obtain $R(D) = 0.302 \pm 0.003$ for $N=2$ and $R(D) = 0.299 \pm 0.004$ for $N=3$, without using the strong 
unitarity constraints. These are all consistent with earlier predictions. 

Also, we predict $R(D^*)$ with the available known corrections at order ${\cal O}(\Lambda_{QCD}/m_{b,c},\alpha_s)$ in the HQET relations between the form factors, 
and we obtain $R(D^*) = 0.259 \pm 0.003$ in the CLN parameterization, while that in the BGL parameterization of the form factors is given 
by $R(D^*) = 0.257 \pm 0.005$. For completeness, we parameterized the unknown corrections in the ratios of the HQET form factors by introducing additional 
factors ($\Delta$s), and fit them from the available data and lattice. After incorporating all the fit results, in the CLN method, we obtain $R(D^*) = 0.259 \pm 0.006$,
while in the BGL method, our best result is $R(D^*) = 0.257 \pm 0.005$.

\section{Acknowledgement}
We would like to thank Paolo Gambino for providing us the complete data file from Belle and having some useful discussions.

\end{document}